\documentclass[aps,prb,twocolumn,superscriptaddress,dvipsnames,longbibliography]{revtex4-2}

\usepackage{amssymb,amsfonts,amsmath} 
\usepackage{graphicx,epsfig,psfrag}
\usepackage{color}
\usepackage{natbib}
\usepackage{url}
\usepackage[breaklinks=true]{hyperref}
\usepackage{mathtools}
\usepackage{subfigure}
\hypersetup{
        colorlinks = true,
        urlcolor = blue,
        citecolor = blue
}
\usepackage{bookmark}
\usepackage{epic}
\usepackage{bbm}
\usepackage{braket}
\usepackage{stmaryrd}
\usepackage{xcolor}
\usepackage{csquotes}
\usepackage{tikz}
\usepackage{extarrows}
\usetikzlibrary{trees}
\usetikzlibrary{decorations.pathmorphing}
\usetikzlibrary{decorations.markings}
\usetikzlibrary{positioning,arrows}
\usetikzlibrary{shapes.arrows}
\usepackage{leftidx}
\usepackage{framed}
\usepackage{tcolorbox}
\usepackage{bm}
\usepackage{physics}
\usepackage{lipsum}
\usepackage{outlines}
\usepackage{longtable}
\usepackage{makecell}
\usepackage{import}
\usepackage{tikz}
\usetikzlibrary{trees}
\usetikzlibrary{decorations.pathmorphing}
\usetikzlibrary{decorations.markings}
\usetikzlibrary{positioning,arrows}
\usetikzlibrary{shapes.arrows}
\usepackage{tkz-tab}

\newcommand{\sign}{\text{sign}}

\newcommand{\rd}[1]{\textcolor{red}{B#1}}
\newcommand{\bl}[1]{\textcolor{blue}{L#1}}

\tikzset{->-/.style={decoration={
  markings,
  mark=at position #1 with {\arrow{>}}},postaction={decorate}}} 

\definecolor{lime}{HTML}{A6CE39}
\DeclareRobustCommand{\orcidicon}{%
    \begin{tikzpicture}
    \draw[lime, fill=lime] (0,0) 
    circle [radius=0.16] 
    node[white] {{\fontfamily{qag}\selectfont \tiny ID}};
    \draw[white, fill=white] (-0.0625,0.095) 
    circle [radius=0.007];
    \end{tikzpicture}
    \hspace{-2mm}
}

\foreach \x in {A, ..., Z}{%
    \expandafter\xdef\csname orcid\x\endcsname{\noexpand\href{https://orcid.org/\csname orcidauthor\x\endcsname}{\noexpand\orcidicon}}
}
\newcommand{\orcid}[1]{\href{https://orcid.org/#1}{\textcolor[HTML]{A6CE39}{\orcidicon}}}

\begin{document} 

\title{Reentrant localisation transitions and anomalous spectral properties in off-diagonal quasiperiodic systems}

\author{Hugo Tabanelli}
\affiliation{\mbox{TCM Group, Cavendish Laboratory, University of Cambridge, Cambridge, CB3 0HE, United Kingdom}}
\affiliation{\mbox{ENS Paris Saclay, 4 Av. des Sciences, 91190 Gif-sur-Yvette, France}}
\author{Claudio Castelnovo \orcid{0000-0003-1752-6343}}
\affiliation{\mbox{TCM Group, Cavendish Laboratory, University of Cambridge, Cambridge, CB3 0HE, United Kingdom}}
\author{Antonio \v{S}trkalj \orcid{0000-0002-9062-6001}}
\email{astrkalj@phy.hr}
\affiliation{\mbox{TCM Group, Cavendish Laboratory, University of Cambridge, Cambridge, CB3 0HE, United Kingdom}}
\affiliation{\mbox{Department of Physics, Faculty of Science, University of Zagreb, Bijeni\v{c}ka c. 32, 10000 Zagreb, Croatia}}

\begin{abstract} 
We investigate the localisation properties of quasiperiodic tight-binding chains with hopping terms modulated by the interpolating Aubry-Andr\'e-Fibonacci (IAAF) function. This off-diagonal IAAF model allows for a smooth and controllable interpolation between two paradigmatic quasiperiodic models: the Aubry-Andr\'e and the Fibonacci model. 
Our analysis shows that the spectrum of this model can be divided into three principal bands, namely, two molecular bands at the edge of the spectrum and one atomic band in the middle, for all values of the interpolating parameter. We reveal that the states in the molecular bands undergo multiple re-entrant localisation transitions, a behaviour previously reported in the diagonal IAAF model. We link the emergence of these re-entrant phenomena to symmetry points of the quasiperiodic modulation and, with that, explain the main ground state properties of the system.
The atomic states in the middle band show no traces of localised phases and remain either extended or critical for any value of the interpolating parameter. Using a renormalisation group approach, adapted from the Fibonacci model, we explain the extended nature of the middle band. These findings expand our knowledge of phase transitions within quasiperiodic systems and highlight the interplay between extended, critical, and localised states.
\end{abstract}
\date{\today} 
\maketitle

%
\section{Introduction}
%
One of the quintessential phenomena in condensed matter physics is Anderson localisation~\cite{Anderson1958, Abrahams2010, Evers2008}, which delineates the quantum confinement of particles such as electrons and photons within randomly disordered systems. Distinct from periodic systems, where the single-particle wavefunctions manifest as extended Bloch waves, disordered media foster conditions where these states can undergo exponential localisation due to destructive interference of scattered waves. Remarkably, within one- and two-dimensional systems, the phenomenon mandates that all single-particle states localise irrespective of the strength of random disorder~\cite{Abrahams1979}. 
The implications of Anderson localisation in such systems extend deeply into their properties, affecting their conductivity and optical transmission~\cite{Lee1985,Akkermans2007,Abrahams2010,Billy2008,Roati2008}.

Another instance of Anderson localisation can be found in quasiperiodic systems~\cite{Shechtman1984,Senechal1995,Suck2013}. They fall, in a sense, between periodic and disordered random systems because, unlike their uncorrelated disordered counterparts, quasiperiodic systems exhibit deterministic patterns, and unlike crystals, they lack translational symmetry yet are not completely random. This unique structural characteristic enables the investigation of localisation phenomena under conditions that diverge from traditional random systems proposed by Anderson~\cite{Anderson1958}, introducing a new paradigm for understanding quantum confinement. The study of localisation in quasiperiodic systems has unveiled a rich tapestry of physical behaviours, including the metal-to-insulator transition in one and two dimensions~\cite{Aubry1980,Jitomirskaya1999,Aulbach2004,Szabo2020,Sbroscia2020,Strkalj2022}, the emergence of critical states~\cite{Kohmoto1983, Ostlund1983, Jitomirskaya1999, Suck2013}, and the fractal nature of the energy spectrum and corresponding eigenmodes~\cite{Kohmoto1984,Kohmoto1986,Kohmoto1987,Niu1986,Mace2016,Jagannathan2021}, which are absent in both periodic and purely random systems.

Two canonical examples of quasiperiodic models are the Aubry-Andr\'e (AA)~\cite{Aubry1980,Jitomirskaya1999} and the Fibonacci model~\cite{Kohmoto1983,Ostlund1983}.
The quasiperiodicity of the AA model comes from cosine modulation of either on-site potential, so-called diagonal modulation, or hopping terms, known as the off-diagonal modulation, that is incommensurate with the underlying periodic lattice. In the Fibonacci model, the modulation is given by two discrete values that appear interchangeably according to the Fibonacci sequence. The two aforementioned models have distinctive localisation properties. Specifically, while the AA model hosts a self-dual localisation transition at a critical potential strength in the case of diagonal modulation or critical-to-extended transition when the quasiperiodic modulation is off-diagonal, the Fibonacci model always has critical wavefunctions.

Despite looking dissimilar at first sight, the two models can be unified in a single one, known as the interpolating Aubry-Andr\'e-Fibonacci model (IAAF)~\cite{Kraus2012}, where a tunable parameter allows for smooth interpolation between the AA and Fibonacci limits.
The diagonal version of the model has been studied in the context of topology~\cite{Kraus2012}, and recently gained much attention for its anomalous localisation properties~\cite{Kraus2014, Strkalj2020, Strkalj2021, Zhai2021, Junmo2023}. More precisely, in Ref.~\onlinecite{Strkalj2020} it was shown that the model hosts re-entrant localisation phenomena as a function of the strength of the modulation and the interpolating parameter.
This peculiar effect inspired further research on the diagonal IAAF model subjected to many-body interactions~\cite{Strkalj2021} and non-Hermicity~\cite{Zhai2021}, as well as the search for other systems showing re-entrant localised phases~\cite{Roy2021,Jiang2021,Wu2021,Padhan2022,Wang2023,Aditya2023,Li2024,Dai2023,Lu2023}.
On the other hand, the off-diagonal IAAF model has been used to investigate topological properties of Fibonacci chains~\cite{Verbin2013}, and to generate topological pumps~\cite{Verbin2015}.
However, its localisation properties remain heretofore unexplored.

In this work, we investigate the localisation properties of the off-diagonal IAAF model, i.e., when the hopping terms follow a quasiperiodic pattern. We show that states living in molecular bands -- namely, closer to the edges of the spectrum -- experience similar multiple re-entrant localisation phenomena as it was found in the diagonal model~\cite{Strkalj2020,Strkalj_thesis}. We extend the previous results, which described the main mechanism responsible for such localisation, and make a direct connection between symmetry points of the quasiperiodic modulation and the localisation centres in different localised phases of the ground state. Furthermore, in each localised phase, we find a sequence over which the ground state wavefunction has the largest support and show that such a sequence grows as the model is tuned towards the Fibonacci limit.
By investigating the middle band states, i.e., the atomic states, we find a contrasting behaviour that has not been reported before. Namely, the states never localise in the thermodynamic limit; in fact, an extended phase exists for all finite values of the interpolating parameter. We elucidate such behaviour and make a connection to a known renormalisation group approach of the Fibonacci model.
Our results extend the knowledge of complex phase transitions between extended, critical, and localised states, and enhance our theoretical understanding of localisation phenomena in quasiperiodic systems. 

The paper is structured as follows. In Sec.~\ref{sec:model}, we introduce the off-diagonal IAAF model and discuss the known localisation properties of its two limiting cases, namely the
off-diagonal Aubry-Andr\'e and Fibonacci models. 
In Sec.~\ref{sec:results}, we present the main results of this paper. We first discuss the main features of the whole spectrum and derive the analytical formula for the extended-to-critical transition. After that, in Sec.~\ref{subsec:gs}, we show that the ground state experiences multiple localisation transitions and we predict the main features of the localised wavefunction, namely its position and width. 
In Sec.~\ref{subsec:middleband}, we turn to the localisation properties of the states located in the middle band. We show that the state in the middle of the spectrum never localises in the thermodynamic limit, but, starting from the AA limit, it experiences a step-wise increase of its inverse participation ratio with universal scaling of the interpolating parameter, until becoming critical in the Fibonacci limit. We explain such behaviour and make a connection to the renormalisation group approach originally developed for the Fibonacci model. 
Lastly, our findings are summarised in Sec.~\ref{sec:discussion_and_conclusions}, and technical details are relegated to the Appendices.

%
\section{Model  \label{sec:model}}
%
\begin{figure}[t!]
\centering
\includegraphics[scale=1]{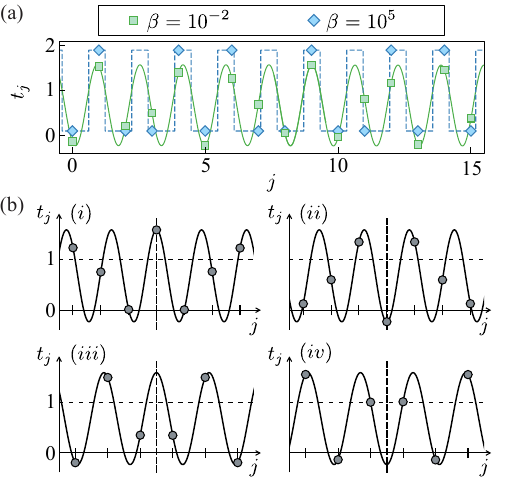}
\caption{
    (a) Spatial distribution of the hopping terms throughout the chain, following the interpolating function in Eq.~\eqref{eq:func_iaaf}. For $\beta \rightarrow 0$, the hopping terms follow the cosine modulation -- whose frequency is irrational -- known as the Aubry-Andr\'e sequence (green line with squares), while in the $\beta \rightarrow \infty$ limit, the sequence is obtained from the continuous set of step functions, in which case it is known as the Fibonacci sequence. 
    (b) The four mirror-symmetry points of the Aubry-Andr\'e modulation function. The ones marked in panels (\textit{i}-\textit{iii}) determine the localisation of the ground state, while the one in panel (\textit{iv}) is crucial for the $E=0$ state. All symmetry points retain their positions for any $\beta$. 
    Note that discrete symbols in all plots denote the bonds, and not the sites of the system.
}
\label{fig:hopping_modulation}
\end{figure} 
We study a quasiperiodic one-dimensional chain with off-diagonal modulation, namely a spatially varying hopping, and uniform (vanishing) onsite potential. The Hamiltonian of the system is~\cite{Kraus2012,Verbin2013,Verbin2015}
\begin{align} \label{eq:ham}
    H = t \sum_{j=0}^{L-1} \underbrace{\left[\vphantom{\sum} 1 - \lambda V_j(\beta) \right]}_{\equiv t_j(\lambda, \beta)} \, \left( c_{j+1}^{\dag}c_j + c_{j}^{\dag}c_{j+1} \right) 
    \, ,  
\end{align}
where without loss of generality we set the constant part of the hopping term to unity, i.e., $t=1$.
Furthermore, in the same equation, $L$ is the number of bonds connecting neighbouring sites, $c^{\dag}_j$, $c_j$ are creation and annihilation operators at site $j$, $\lambda$ is the amplitude of the hopping modulation, and $V_j(\beta)$ is the interpolating function given by
\begin{align} \label{eq:func_iaaf}
    V_j(\beta) = - \frac{ \tanh \Big[ \beta [\cos(2\pi b j + \varphi) - \cos(\pi b)] \Big]}{\tanh(\beta)} \, . 
\end{align}
The frequency of the function above is irrational -- set by the inverse of the golden mean $b = 2/(1+\sqrt{5})$ -- and, therefore, the system is quasiperiodic. Furthermore, as already discussed in prior works~\cite{Aubry1980,Strkalj2020,Strkalj_thesis,Strkalj2021}, the phase $\varphi$ does not affect the localisation properties of the system and, without loss of generality, we set it to $\varphi = 3 \pi b$.
For convenience, we use the following abbreviation $t_j(\lambda, \beta) \equiv 1 - \lambda V_j(\beta)$.
Notice that in our convention, the bond indexed by $j$ is between sites $j$ and $j+1$.

The interpolating function in Eq.~\eqref{eq:func_iaaf} has two well-known limits. In the limit of $\beta = 0$, the function becomes $V_j =  - \cos(2\pi b j + \varphi) + \cos(\pi b)$, as shown in Fig.~\ref{fig:hopping_modulation}(a) by a green line, leading to the off-diagonal Aubry-Andr\'e (AA) model~\cite{Thouless1983,Han1994,Chang1997,Takada2004,Liu2015,Tang2021,Cai2022,He2022,Li2023}. 
The constant shift of $\cos(\pi b)$ in Eq.~\eqref{eq:func_iaaf} ensures that starting from the AA limit, one ends in the correct Fibonacci limit.
Upon increasing $\beta$, the modulation~\eqref{eq:func_iaaf} becomes steeper, until it reaches the shape of a repeating step-function, 
$V_j = -2 \big( \lfloor(1+b)^{-1} (j+2)\rfloor-\lfloor(1+b)^{-1} (j+1)\rfloor \big)+1 = \pm 1$~\cite{Verbin2015}, 
in the limit $\beta \rightarrow \infty$, as shown in Fig.~\ref{fig:hopping_modulation}(a) by a dashed blue line. For discrete $j$, the aforementioned step-function yields the Fibonacci sequence, which is made of two letters (say, $W$ and $S$) that interchange according to the Fibonacci rule. 
An alternative way of obtaining the Fibonacci sequence, denoted by ${\omega_n}$, is iterative, using the relation $\omega_n = \omega_{n-1} \omega_{n-2}$ with the initial condition $\omega_0 = S$ and $\omega_1= W$. For example, the next four sequences $\{ \omega_2, \omega_3, \omega_4, \omega_5 \}$ are $\{ WS, WSW, WSWWS, WSWWSWSW \}$, and the number of letters, denoted by  $L$, in each aforementioned sequence is given by the Fibonacci number $L=\{ F_2, F_3, F_4, F_5 \}$ = \{ 2, 3, 5, 8 \}. The ratio between the numbers of $S$ and $W$ letters in a sequence of length $L=F_n$ is $F_{n-2}/F_{n-1}$~\cite{Mace_thesis2017}, which approaches the value $b$ when $n \rightarrow \infty$. 
Lastly, as it is clear from Eqs.~\eqref{eq:ham} and~\eqref{eq:func_iaaf} and from Fig.~\ref{fig:hopping_modulation}(a), the values of the hopping terms with $t_j > 1$ ($t_j < 1$) 
tend to $t_j = 1+\abs{\lambda}$ ($t_j = 1-\abs{\lambda}$) when $\beta \rightarrow \infty$. 
The continuity of the $\tanh$ as a function of $\beta$ guarantees that the ratio between the number of $t_j > 1$ and $t_j < 1$ stays the same for all $\beta$, and it is precisely equal to $F_{n-2}/F_{n-1}$. 

The hopping function $t_j$ given by Eqs.~\eqref{eq:ham} and~\eqref{eq:func_iaaf} has an important feature that we shall use later when discussing the localisation properties. Namely, it contains four mirror-symmetry points, shown in Fig.~\ref{fig:hopping_modulation}(b). 
Two mirror-symmetry points are located at a bond, i.e., between two neighbouring sites on the chain, and two are at a site, as illustrated by the vertical dashed lines in the upper and lower panels in Fig.~\ref{fig:hopping_modulation}(b), respectively. The former are given by the positions of the absolute maximum and minimum of $t_j$, while the latter are placed at a site whose left and right hopping are degenerate.

%
\subsection{Localisation properties in the Aubry-Andr\'{e} limit 
\label{subsec:AA_limit}}
%
Before presenting our novel results on the off-diagonal IAAF model, we briefly summarise the properties of its two limiting cases, namely the off-diagonal Aubry-Andr\'{e} and Fibonacci models. 
The former has been studied as a generalised AA model, with both diagonal and off-diagonal cosine-modulated terms~\cite{Hatsugai1990,Han1994,Chang1997,Takada2004,Liu2015,He2022}. By analysing the bandwidth of the model~\cite{Thouless1983}, it was shown that the phase diagram does not exhibit localised phases in the limit of purely off-diagonal modulation~\cite{Thouless1983,Han1994,Chang1997,Takada2004,Liu2015,Tang2021,Cai2022,He2022,Li2023}, in stark contrast to a purely diagonal model~\cite{Aubry1980,Jitomirskaya1999}. In fact, the off-diagonal AA model undergoes a transition from an extended to a critical phase -- experienced by all states simultaneously -- which occurs when the weakest hopping in the chain vanishes~\cite{Thouless1983}, i.e., when $\min[t_j]=0$.
In other words, the system is in a critical phase when $\min[t_j(\lambda)] \leq 0$, cf. Fig.~\ref{fig:spectra}(a).

%
\subsection{Localisation properties in the Fibonacci limit 
\label{subsec:Fibonacci_limit}}
The Fibonacci model is known not to undergo a localisation transition either in the diagonal~\cite{Kohmoto1983,Ostlund1983,Kohmoto1984} or in the off-diagonal case, for any values of $\lambda \neq \pm 1$. Instead, it exhibits a critical spectrum~\cite{Jagannathan2021} which forms a Cantor set with zero Lebesgue measure, i.e., it is purely singular and continuous, with wavefunctions being neither localised nor fully extended~\cite{Kohmoto1986,Niu1986,Halsey1986,Kohmoto1987,Zheng1987,Niu1990,Damanik2008,Piechon1995,Mace2016}. The wavefunctions have a critical scaling, which can be seen in the fractal dimension or inverse participation ratio, as we discuss later.

%
\subsubsection{Renormalisation group approach for the Fibonacci model}
%
\begin{figure}[t!]
    \centering    \includegraphics{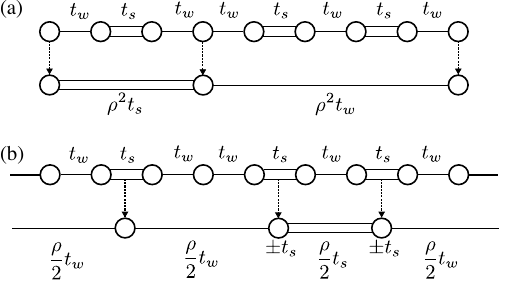}
    \caption{Illustration of the (a) atomic and (b) molecular renormalisation group (RG) step in the Fibonacci chain. Here, the fifth approximant, with $L=F_5$, is mapped to the second, $L=F_2$, in (a) and the third, $L=F_3$, in (b). Smaller chains have renormalised hopping terms that are scaled by a factor $\rho^2$ in the atomic and $\rho/2$ in the molecular RG.}
    \label{fig:RG_Fib}
\end{figure} 
A useful tool to understand the critical spectrum and wavefunctions of the off-diagonal Fibonacci model is the real space renormalisation group (RG) scheme~\cite{Niu1986,Niu1990,Piechon1995,Mace2016,Jagannathan2021}.
As already mentioned, the off-diagonal Fibonacci model is obtained by setting $\beta \rightarrow \infty$ in Eq.~\eqref{eq:func_iaaf}, in which case the system contains only two types of hopping terms: weak, $t_w = 1 - |\lambda|$, and strong, $t_s = 1 + |\lambda|$, illustrated in Fig.~\ref{fig:hopping_modulation}. 
Therefore, the properties of the spectrum reduce to a single parameter, $\rho = t_w/t_s$. 

Following the Fibonacci sequence, sites can sit in between two weak bonds (`W', in our two-letter example above) or between one weak (`W') and one strong (`S') bond. In the $\rho = 0$ limit, or equivalently, when $\lambda = \pm 1$, the weak hopping terms vanish and the chain disconnects into isolated sites and pairs of sites sharing a strong bond. The former can be treated as `atoms' decoupled from the rest of the chain, while the latter form `molecular' states extending over two sites. Therefore, one can refer to sites that are surrounded by two $t_w$ hopping terms as atomic sites, while the other ones are called molecular sites.
The spectrum of a chain with $t_w=0$ contains three energy levels, one atomic $E=0$ level and two molecular $E=\pm t_s$ levels -- all three being highly degenerate. For a chain containing $F_n$ hopping terms, the degeneracy of the atomic level is $F_{n-3}$ and of the molecular ones is $F_{n-2}$, for each $\pm t_s$ level.

For finite but small weak hopping $t_w \ll t_s$, when $\rho \ll 1$, one can apply the RG scheme~\cite{Niu1986,Niu1990,Piechon1995,Mace2016,Jagannathan2021} separately for atomic and molecular states, thanks to the fact that atomic and molecular energy levels do not mix strongly. Schematics of the RG scheme are shown in Fig.~\ref{fig:RG_Fib}.
After the atomic RG step, performed on a chain with $F_n$ sites, the number of atomic sites in the new chain reduces to $F_{n-3}$ and renormalised weak and strong hopping terms become $t_{w/s}' = \rho^2 t_{w/s}$.
On the other hand, after the molecular RG step, the chain size reduces to $F_{n-2}$ with renormalised hopping terms $t_{w/s}'' = \frac{\rho}{2} t_{w/s}$, and additional on-site energy $\pm t_s$. 
Generally, to lowest order in $\rho$, the Hamiltonian of the $n$-th approximant $F_n$ can be decoupled into a direct sum of molecular (i.e., bonding and antibonding) and atomic part: 
{\small
\begin{align}   \label{eq:ham_rec_fib}
    H^{(n)} = \left( \frac{\rho}{2} H^{(n-2)} - t_s\right) \oplus \left( \rho^2 H^{(n-3)} \right) 
    \oplus \left( \frac{\rho}{2} H^{(n-2)} + t_s \right) \, .
\end{align}
}
The recursive relation above means that the spectrum is a Cantor set where the energy bands decompose at each RG step into three subbands. After each RG step, one can uniquely label all energy subclusters by their renormalisation path, and therefore, each state in the finite chain has a unique renormalisation path.
Using the same RG approach, it is easy to show the critical nature of the wavefunctions, see Refs.~\onlinecite{Niu1986,Niu1990,Piechon1995,Mace2016,Jagannathan2021}.

%
\section{Results for the interpolating model   \label{sec:results}}
%
\begin{figure}[t]
    \centering
    \includegraphics[scale=1]{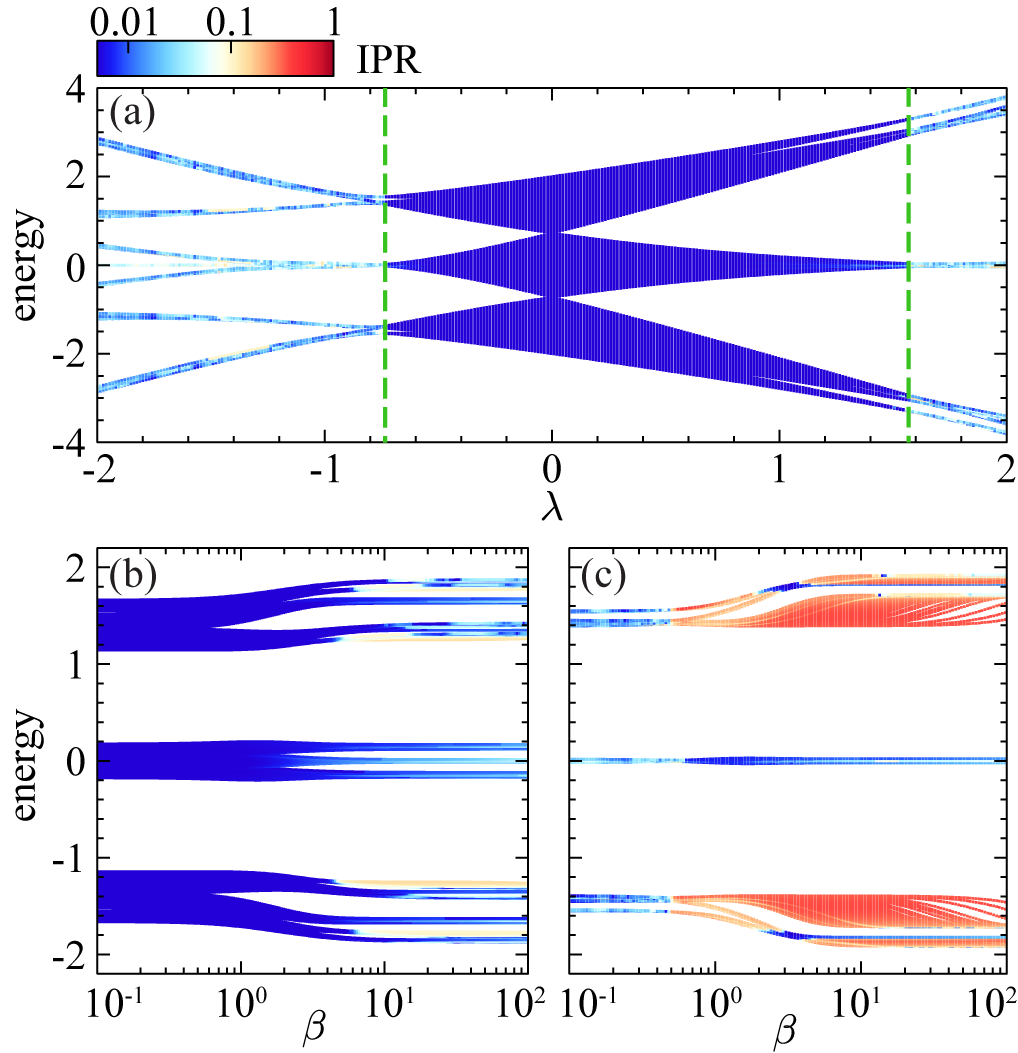}
    \caption{Localisation properties of the spectrum of the off-diagonal IAAF model.
    (a) IPR, Eq.~\eqref{eq:IPR}, of the eigenstates of Eq.~\eqref{eq:ham} as a function of $\lambda$ close to the AA limit, namely for $\beta = 10^{-5}$. 
    Vertical dashed lines mark analytical predictions for the transition from the extended phase at small $|\lambda|$ to the critical phase at large $|\lambda|$, see Eq.~\eqref{eq:ldc}. 
    (b),(c) IPR of the spectrum as a function of $\beta$ for constant $\lambda = -0.5$ and $\lambda=-0.8$, respectively. In (c), the localisation of the spectrum is highly non-homogeneous, with some states in the lower and upper bands experiencing multiple localisation-delocalisation transitions.
    For all plots, we used chains with $L=F_{14}=610$ sites and imposed periodic boundary conditions. 
    }
    \label{fig:spectra}
\end{figure}
After a brief summary of the known results for the off-diagonal AA and Fibonacci models, we turn to the analysis of the off-diagonal IAAF model defined in Eqs.~\eqref{eq:ham} and~\eqref{eq:func_iaaf}.
To gain insight into the localisation properties of the model, we first study the moments of the normalised energy eigenstates $\Psi_j(E_n)$, defined as
\begin{align} \label{eq:IPR}
    \chi_q(E_n) = \sum_{j=1}^{L} |\Psi_j(E_n)|^{2q} 
    \, ,
\end{align}
where $\chi_1=1$ follows from the normalisation condition and $\chi_2$ is the inverse participation ratio (IPR).
The IPR is one of the simplest measures of localisation. It is inversely proportional to the system size for eigenstates extended across the whole system, $\mathrm{IPR} \sim 1/L$, and approaches zero in the thermodynamic limit. On the other hand, for states that are exponentially localised across $l$ sites, $\mathrm{IPR} \sim 1/l$ remains finite in the thermodynamic limit. 

We start with the small-$\beta$ limit of Eq.~\eqref{eq:func_iaaf}, i.e., the AA limit, and show its spectral properties in Fig.~\ref{fig:spectra}(a). The IPR of each eigenstate shows homogeneous localisation throughout the spectrum at constant $\lambda$. For small $|\lambda|$, the IPR of each eigenstate is low and suggests an extended state, while at larger $|\lambda|$ the intermediate values of the IPR relate to the critical nature of the eigenstates~\cite{Thouless1983,Han1994,Chang1997,Takada2004,Liu2015,Tang2021,Cai2022,He2022,Li2023}. 
The extended phase in Fig.~\ref{fig:spectra}(a) is bounded by
\begin{align}   \label{eq:ldc}
    \lambda^{\pm}_{\rm c}(\beta)= \frac{\tanh(\beta)}{\tanh\{\beta [ \cos(\pi b) \pm 1]\}} \, ,
\end{align}
which is obtained by finding the critical $\lambda$ where the minimum of the continuous function $t(\lambda, \beta)$ is equal to zero, i.e., by solving $\min[t(\lambda^{\pm}_{\rm c}, \beta)] = 0$. 
For vanishing $\beta$, the above expression becomes $\lambda_{\rm c}^{\pm} = \left[ \cos(\pi b) \pm 1 \right]^{-1}$; as it can be seen from the dashed lines that mark $\lambda_{\rm c}^{\pm}$ in Fig.~\ref{fig:spectra}(a), this limit accurately describes the bounds of the numerically calculated extended phase of the model. 

For finite but small $\beta$, the localisation properties remain the same as in the AA limit, see Figs.~\ref{fig:spectra}(b) and (c), where two representative values of $\lambda$ are used. In Fig.~\ref{fig:spectra}(b), $|\lambda|<\lambda_{\rm c}$ and all eigenstates are extended, while in Fig.~\ref{fig:spectra}(c) they are critical since $|\lambda|>\lambda_{\rm c}$. 

As $\beta$ is increased, the spectra of both cases shown in Figs.~\ref{fig:spectra}(b) and (c) exhibit inhomogeneous behaviour of the IPR as a function of both energy and $\beta$. For small $|\lambda|$ in the extended region of the AA limit, shown in Fig.~\ref{fig:spectra}(b), the IPR of all eigenstates seems to increase by tuning $\beta$, although not uniformly across the spectrum. This behaviour becomes more pronounced for $\lambda < \lambda_{\rm c}$, in the critical phase of the AA limit, as shown in Fig.~\ref{fig:spectra}(c). It is evident that different eigenstates do not share the same localisation properties; while the IPR reduces for the states in the atomic band, it increases to a value close to one for the states in the molecular bands. Even more surprisingly, the IPR of molecular bands exhibits non-monotonic behaviour as a function of $\beta$, indicating that the states experience multiple transitions from localised to extended or critical phases. 
Similar inhomogeneous reentrant localisation transitions were observed and discussed in the diagonal IAAF model by one of the authors in Refs.~\onlinecite{Strkalj2020, Strkalj2021}.

%
\subsection{Properties of the ground state  \label{subsec:gs}}
%
In this Subsection, we focus on the molecular bands and we explain the non-monotonic localisation behaviour as a function of $\beta$. We choose the ground state as a representative state of the molecular bands.

%
\subsubsection{Phase diagram \label{subsec:phase_diagram}}
%
\begin{figure*}[ht!]
\centering
    \includegraphics[scale=1]{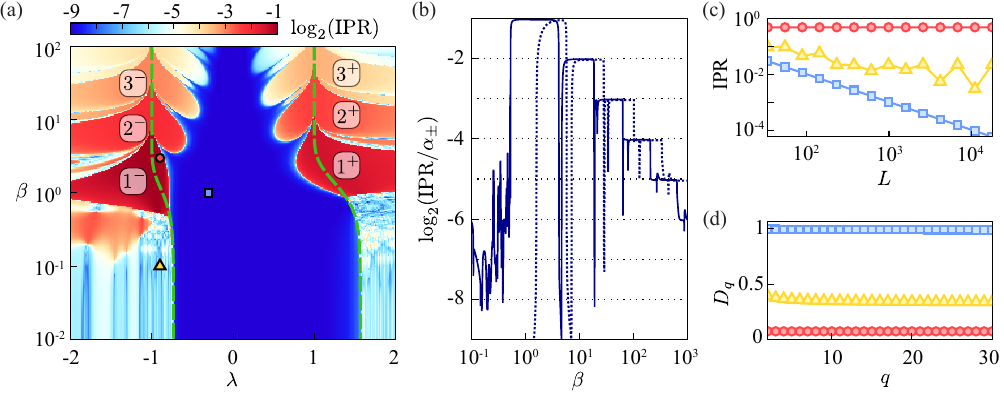}
    \caption{Localisation properties of the ground state. 
    (a) IPR of the ground state as a function of $\lambda$ and $\beta$ for $L=F_{14}=610$. The green dashed lines are analytical predictions for $\lambda^{\pm}_c$ given by Eq.~\eqref{eq:ldc}. Three geometric markers have been placed to show the values of parameters $\lambda$ and $\beta$ used in (c) and (d): a yellow triangle, a light-blue square, and a red circle. 
    (b) IPR of the ground state as a function of $\beta$ only, for selected values of $\lambda$. The IPR is scaled by a constant $\alpha_{\pm}$, see Eq.~\eqref{eq:IPR_GS}, to emphasise the same localisation mechanism for both positive and negative $\lambda$.
    The solid line is for $\lambda=-0.9$, while the dotted line is for $\lambda=0.9$. Here we used a system with $L=F_{16}=1597$.
    (c) IPR as a function of system size for the choice of parameters indicated by the markers in panel (a). We show only system sizes equal to a Fibonacci number, as required to remain consistent in the $\beta \to \infty$ limit. 
    (d) Fractal dimensions $D_q$, defined in Eq.~\eqref{eq:fractal_dimension}, as a function of $q$ for the same three points in parameter space identified by the symbols in (a). For this plot, we used $L= F_{21} = 17711$.}
    \label{fig:phase_diagram_GS}
\end{figure*} 
We first calculate numerically the IPR of the ground state wavefunction and show the results in Figs.~\ref{fig:phase_diagram_GS}(a-c). The resulting localisation phase diagram as a function of $\beta$ and $\lambda$ is shown in Fig.~\ref{fig:phase_diagram_GS}(a). 
The limiting case of $\beta \rightarrow 0$ gives the expected extended-to-critical phase transitions, and the position of the three regions remains almost unchanged up to $\beta \approx 1$. The analytical prediction for the transition given by Eq.~\eqref{eq:ldc}, and marked by dashed lines in Fig.~\ref{fig:phase_diagram_GS}(a), is also found to be in good agreement with the numerics up to the appearance of localised (red) regions around $\beta \approx 1$. 

By increasing $\beta$ further, the extended region shrinks and lobes with larger IPR appear for both positive and negative $\lambda$, indicating the presence of localised phases. The value of IPR is constant inside each localised lobe, and it reduces with increasing $\beta$. Surprisingly, the IPR value drops by a factor of $2$ from one localised lobe to the next as $\beta$ is increased, as can be seen from the colour of the lobes and the $\log_2$ scale in Fig.~\ref{fig:phase_diagram_GS}(a), and from the sharp steps in Fig.~\ref{fig:phase_diagram_GS}(b). 
If we enumerate the localised phases with $p^{\pm}$, where the superscript denotes positive / negative $\lambda$, from what we observe in Figs.~\ref{fig:phase_diagram_GS}(a) and (b) it follows that 
\begin{align}   \label{eq:IPR_GS}
    {\rm IPR}^{\pm}_p = \alpha_{\pm} 2^{-p}
    \, , 
\end{align}
with $\alpha_-=1$ and $\alpha_+ = 3/4$. The origin of these two constants will be discussed at the end of the next Subsection. 

To further confirm that the ground state is indeed localised in the phases labelled $p^{\pm}$, we study how the IPR scales with system size while keeping $\lambda$ and $\beta$ constant in Fig.~\ref{fig:phase_diagram_GS}(c). The IPR is inversely proportional to the chain length with an exponent equal to $1$ in the extended phase (blue squares) and shows a slower decay, with an exponent smaller than $1$, in the critical phase (yellow triangles). Inside the localised phases (red circles), the IPR remains constant with a value close to $1$. Additional evidence of localisation is given in Fig.~\ref{fig:phase_diagram_GS}(d), where we show the fractal dimensions~\cite{Evers2008} defined as
\begin{align}   \label{eq:fractal_dimension}
    D_q = \frac{1}{1-q} \frac{\ln {\chi}_q}{\ln L}
    \, .
\end{align}
In the thermodynamic limit, $D_q=0$ for localised wavefunctions and $D_q=1$ for extended ones, while for critical wavefunctions it acquires a nontrivial functional dependence on $q$. In Fig.~\ref{fig:phase_diagram_GS}(d), we show the fractal dimensions for the three values of the parameters marked in the phase diagram in Fig.~\ref{fig:phase_diagram_GS}(a). The results for $D_q$ are consistent with our conclusions from Fig.~\ref{fig:phase_diagram_GS}(c). 

%
\subsubsection{Localisation mechanism}
%
\begin{figure}[h]
    \centering
    \includegraphics[scale=1]{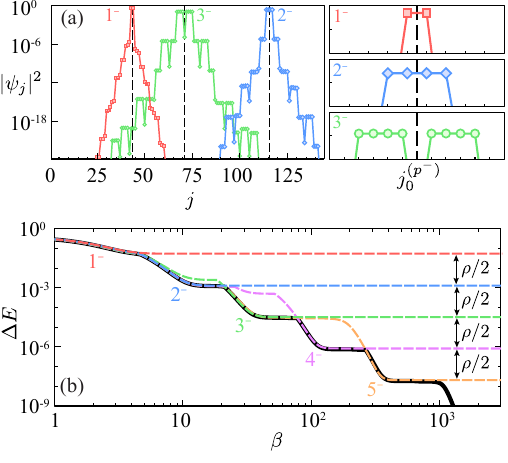}
    \caption{
    (a) Ground state density, $|\psi_j(E_{\rm G})|^2$, as a function of site index $j$ for the localised phases $1^-$, $2^-$, $3^-$ (in red, blue, and green, respectively), with insets showing the behaviour near the maxima. 
    Panel (b) shows the ground state energy, plotted as $\Delta E = E_{p^-} - E_{6^-}$ after subtracting the constant plateau of $E_{6^-}$, as a function of $\beta$. The black line is obtained by exact diagonalisation of the whole chain with $L = F_{17} = 2584$, and the coloured dashed lines correspond to the energies $\Delta E_{p^-}$ obtained by exact diagonalisation of truncated chains, as described in the main text.
    In both plots, we used $\lambda=-0.9$. 
    }
    \label{fig:loc}
\end{figure} 
Let us now study the mechanism responsible for the non-monotonic localisation phase diagram and the stepwise behaviour of the IPR. Firstly, we focus on $\lambda<0$ and start by discussing the $p=1^{-}$ phase. As suggested by the IPR, and confirmed by the plot of the particle density in the ground state, $|\psi(E_G)|^2$ in Fig.~\ref{fig:loc}(a), the state is exponentially localised on two neighbouring sites. One possible reason for this two-site localisation is that it corresponds locally to the case where a single strong hopping is surrounded by much weaker ones. Such a situation occurs at the bond where the function $t_j$ is maximal, i.e., when 
\begin{align}
    t_{j_0^{(1-)}} =  1 + \lambda \frac{\tanh\{\beta [-1 - \cos(\pi b)]\}}{\tanh(\beta)} \, , 
\end{align}
with $j_0^{(1-)}$ marking the location of the strongest bond.
This corresponds to the first symmetry point of the modulated hopping function $t_j$, illustrated in Fig.~\ref{fig:hopping_modulation}(b) case (\textit{i}). Moreover, the period of modulation $1/b$, which is slightly larger than the lattice spacing, guarantees that the first few hopping terms around the maximum are smaller than unity, therefore corresponding to weak hopping terms in the Fibonacci limit. Consequently, increasing $\beta$ further enhances the difference between the aforementioned maximum and its neighbours until the energy of the molecular state -- the one exponentially localised on the two sites sharing the maximal hopping value -- becomes lower than all other energies of the system, leading to the ground state localisation observed in the $1^{-}$ phase. This is confirmed indeed when we compare the position of the maximum of the ground state density with the analytical result for the position of the maximal bond $j_0^{(1^-)}$, as shown in Fig.~\ref{fig:loc}(a).
In App.~\ref{app:localisation_and_symmetry_points}, we present the analytical calculation of the position of the first symmetry point in the infinite chain, and we further confirm that the localisation in the $1^-$ phase can be accurately predicted for any $\varphi$. 

Assuming (as observed) that the ground state is strongly localised on two sites, which is progressively more accurate in the limit $\lambda \to -1$, we can propose to estimate its energy analytically:  
\begin{align}
    E_{1^-}(\beta) &= 
    -t_{j_0^{(1-)}}
    \nonumber \\ 
    &= - 1 - \lambda \frac{\tanh\{\beta [-1 - \cos(\pi b)]\}}{\tanh(\beta)} 
\, .
\label{eq:E1}
\end{align}
A comparison between the ground state energy obtained from the numerical method of exact diagonalisation, as a function of $\beta$ for constant $\lambda = -0.9$, and the analytical expression above is shown in Fig.~\ref{fig:loc}(b), demonstrating good agreement in the range of $\beta$ that corresponds to the $1^-$ phase, illustrated in Fig.~\ref{fig:phase_diagram_GS}(a) and (b). 

In the next localised phase, namely the $p=2^-$ phase that takes place for larger values of $\beta$, the state appears to be exponentially localised on four sites -- as indicated by the value of the second plateau in the IPR in Fig.~\ref{fig:phase_diagram_GS}(b) and by the density profile shown in Fig.~\ref{fig:loc}(a). 
Correspondingly, three hopping terms appear to host most of the wavefunction or, more precisely, most of the density. 
Interestingly, the state has a mirror symmetry around the middle bond. 
The only way to satisfy this is to impose a degeneracy condition $t_{j_0^{(2-)}-1} = t_{j_0^{(2-)}+1}$, with $j_0^{(2-)}$ being the position of the middle bond, and requiring that $t_{j_0^{2-} \pm 1} > 1$. 
Note that the mirror symmetry around the bond, as well as the condition that the two neighbouring bonds host strong hopping terms in the Fibonacci limit, corresponds precisely to the second symmetry point of the hopping modulation function in Fig.~\ref{fig:hopping_modulation}(b), case (\textit{ii}). 
It follows that the three hopping terms are given by:
\begin{align} \label{eq:t_p2}
    t_{j_0^{(2-)} \pm 1} &= 1 + \lambda \frac{\tanh\{\beta [\cos(2 \pi b) - \cos(\pi b)]\}}{\tanh(\beta)}
\, , \nonumber \\
    t_{j_0^{(2-)}} &= 1 + \lambda \frac{\tanh\{\beta [1 - \cos(\pi b)]\}}{\tanh(\beta)}
\, . 
\end{align}

To confirm this scenario, we numerically find the two next-nearest-neighbour bonds in a finite chain that are closest in value, minimising $\abs{t_{j_0^{2-}-1} - t_{j_0^{2-}+1}}$, and we compare the position $j_0^{(2-)}$ with the density profile of the ground state in the $2^-$ phase, see Fig.~\ref{fig:loc}(a). The predicted localisation point matches the mirror symmetry point of the ground-state density $|\psi(E_G)|^2$.
Similarly to the case of the $1^-$ phase, we calculate the ground state energy by assuming complete (compact~\cite{Kirkpatrick1971,Sutherland1986}) localisation on the four sites connected by the three hopping terms in Eq.~\eqref{eq:t_p2}. The lowest energy is given by
\begin{align} \label{eq:E2}
        E_{2^-} = \frac{1}{2} \left( -\abs{t_{j_0^{(2-)}}} - \sqrt{4 \, t_{j_0^{(2-)} \pm 1}^2+ t_{j_0^{(2-)}}^2} \right)
\, , 
\end{align}
which we compare with the numerical result for a large system in Fig.~\ref{fig:loc}(b). Once again, we find good agreement in the range of $\beta$ that corresponds to the $2^-$ phase, illustrated in Fig.~\ref{fig:phase_diagram_GS}(a) and (b).

When we plot in Fig.~\ref{fig:loc} the density profile in the $3^-$ phase, we observe similar features as for the density profiles in the $1^-$ and $2^-$ phases: the largest support of the density is located on a few lattice sites while away from these the density decays exponentially; moreover, the density has mirror symmetry about its middle point, which in this case coincides with the third symmetry point of the IAAF modulation function in Fig.~\ref{fig:hopping_modulation}(b), case (\textit{iii}). From these considerations, an additional feature becomes apparent: the local configuration of sites over which the wavefunction has the largest support in the $3^-$ phase can be seen as made of two copies of the $2^-$ configuration, separated by one bridging site, or equivalently, two bridging hopping terms. 
In hindsight, we could have noticed that the same holds for the local hopping configuration underpinning the $2^-$ phase, being made of two copies of the local configuration of the $1^-$ phase. 

The origin of such a recursive structure is reminiscent of the one observed in Ref.~\onlinecite{Strkalj2020}. Namely, upon increasing the parameter $\beta$, the molecular states are pushed to the edges of the spectrum where they sequentially hybridise and form progressive sequences of lowest (and highest) energy states. Correspondingly, the ground state in each $p$ phase results from the hybridisation of two local configurations of the $(p-1)$ phase.

This can be put in more formal terms by introducing the concept of localisation sequence $L_p$, defined in the Fibonacci limit ($\beta \rightarrow \infty$). It consists of the sequence of weak ($W$) and strong ($S$) hopping terms (that, for finite $\beta$, have values $t_j<1$ and $t_j>1$, respectively) corresponding to the bonds that span the leading support of the ground state density, as discussed above. 
While several different sequences of $t_j$ hopping terms at finite $\beta$ correspond to the same localisation sequence $L_p$, we shall see in the following that the latter contains in fact the essential information needed to understand the localised nature of the corresponding ground state wavefunction. 

Using the definition of $L_p$ and the observations from the previous paragraphs, we can propose the following recurrence relation: $L_p = L_{p-1} B_p L_{p-1}$, where $B_p$ is the bridging sequence made of weak and strong hopping terms that links the two copies of $L_{p-1}$, with a mirror symmetry point in its middle. 
Based on the insets in Fig.~\ref{fig:loc}(a), we obtain the first three localisation sequences: $L_{1} = S$, $L_2 = SWS$ and $L_3 = SWSWWSWS$, from which it follows that $B_1=\varnothing$, $B_2=W$ and $B_3=WW$. 
The full set of localised sequences of the ground states in the phases $p = {1^-, ..., 5^-}$ are shown in Table~\ref{tab:loc}. 
\begin{table*}
    \begin{tabular}{|c|l|l|}
    \hline 
    \multicolumn{1}{|c|}{\textbf{\makecell{Phase Number \\ $p$}}} 
    & \multicolumn{1}{c|}{\textbf{\makecell{Bridge Sequence \\ $B_p$}}} 
    & \multicolumn{1}{c|}{\textbf{\makecell{Localised Sequence \\ $L_p$}}} \\ 
    \hline 
    $1^{-}$  & $\rd{_1} = \varnothing$ & $\bl{_1} = S$ \\ \hline
    $2^{-}$ & $\rd{_2} = W$ & $\bl{_2} = SWS = \bl{_1}\rd{_2}\bl{_1}$ \\ \hline
    $3^{-}$ & $\rd{_3} = WW$ & $\bl{_3} = SWSWWSWS = \bl{_2}\rd{_3}\bl{_2}$ \\ \hline
    $4^{-}$ & $\rd{_4} = WWSWW = \rd{_3}\bl{_1}\rd{_3}$ & $\bl{_4} = ... = \bl{_3}\rd{_4}\bl{_3}$\\ \hline
    $5^{-}$ & $\rd{_5} = WWSWWSWSWWSWW = \rd{_4}\bl{_2}\rd{_4}$ & $\bl{_5} = ... = \bl{_4}\rd{_5}\bl{_4}$\\ \hline
    $p^{-}$ & $\rd{_p} = \rd{_{p-1}}\bl{_{p-3}}\rd{_{p-1}}$ & $\bl{_p} = \bl{_{p-1}}\rd{_p}\bl{_{p-1}}$ \\\hline
    \end{tabular}
    \caption{Localisation sequences, bridging sequences, and recurrence relations for the ground states in the phases $p = {1^-, ..., 5^-}$. The length of the localised sequence $L_p$ is $F_{2p-1}$, while the length of the bridging sequence is $\delta_p = F_{2p-4}$, see Eq.~\eqref{eq:delta_p}.} 
    \label{tab:loc}
\end{table*}
From this, we conjecture the following recursive relation for the bridging sequences: $B_p = B_{p-1} L_{p-3} B_{p-1}$, which indeed respects the mirror symmetry around the middle point. It now follows that the localisation sequence is a finite Fibonacci word whose length is given by the Fibonacci number $F_{2p-1}$. Moreover, it follows that the two consecutive sequences $L_p$ and $L_{p-1}$ are related by a single molecular RG step, described in Sec.~\ref{subsec:Fibonacci_limit}, see also Eq.~\eqref{eq:ham_rec_fib}. In the Fibonacci limit, the energy scaling of a molecular part of the Hamiltonian after applying the molecular RG step is given by a factor $\rho/2$ (where $\rho = t_w/t_s$), which explains the equidistant energy plateaus in Fig.~\ref{fig:loc}(b).

Combining the knowledge of $L_p$ with the fact that the ground state profile is mirror symmetric, i.e., imposing a degeneracy condition on the hopping terms on the left and right side from the symmetry point, allows us to uniquely determine the sequence of hopping terms $t_j$ at finite $\beta$ over which the ground state localises in the phase $p$. 
In other words, the symmetry points in Fig.~\ref{fig:hopping_modulation}(b) give the positions of the highest density point of the ground state wavefunctions, and $L_p$ gives the number of bonds over which each wavefunction localises.  
To find such a sequence of $t_j$ we proceed as follows. We first concentrate on the two hopping terms in the sequence that lie at the closest ends of the $L_{p-1}$ subsequences, and we search for the central $j^{(p^-)}$ that respects the mirror symmetry:
\begin{align}   \label{eq:degeneracy_p}
   t_{\big\lfloor j_0^{(p)} - \frac{\delta_p+1}{2} \big\rfloor } = t_{\big\lfloor j_0^{(p)} + \frac{\delta_p+1}{2} \big\rfloor }
\, ,
\end{align} 
where $\delta_p$ is the distance between the subsequences $L_{p-1}$ (which coincides with the length of the bridging sequence $B_p$).  
From the recurrence relation for $L_p$, and knowing that the length of $L_p$ is $F_{2p-1}$, it is easy to derive the following expression
\begin{align}   \label{eq:delta_p}
    \delta_p = F_{2p-1} - 2 F_{2p-3} = F_{2p-4}\, .
\end{align}
Combining Eqs.~\eqref{eq:degeneracy_p} and~\eqref{eq:delta_p} with the location of $j^{(p^-)}$ obtained from Eq.~\eqref{eq:degeneracy_p}, i.e., from the symmetry points in Fig.~\ref{fig:hopping_modulation}(b) and also Fig.~\ref{fig:loc}(a), we obtain the sequence of hopping terms in $L_p$: 
\begin{align} \label{eq:hoppings_loc_sequence}
    t_j \in L_p \:
    \mathrm{if} \: j \in \Big[& 
    \Big\lfloor j_0^{(p)} - \frac{\delta_p+1}{2} \Big\rfloor
    -(F_{2p-3}-1), \nonumber\\
    &\Big\lfloor j_0^{(p)} + \frac{\delta_p+1}{2} \Big\rfloor
    +(F_{2p-3}-1)
    \Big] \, ,
\end{align}
where $F_{2p-3}$ is the size of each subsequence $L_{p-1}$.
Lastly, note that for finite size chains, the condition~\eqref{eq:degeneracy_p} is not exact, namely the degeneracy is not perfect. However, the condition can be easily modified by searching for a pair of hopping terms that minimise $\abs{t_{\lfloor j_0^{(p)} - \frac{\delta_p+1}{2} \rfloor} - t_{\big\lfloor j_0^{(p)} + \frac{\delta_p+1}{2} \big\rfloor}}$ instead.

Finally, we compare results from the exact diagonalisation of the truncated chain with $\widetilde{L_p}$ bonds with the numerical results for the full chain with $L=F_n \gg \widetilde{L_p}$. The $\widetilde{L_p} > L_p$ is an extended sequence that contains $L_p$ and shares the same $j_0^{(p^-)}$ with it, see the definition of $\widetilde{L_p}$ and the discussion about why we use it in App.~\ref{app:GS_energy_comparison}. 
In Fig.~\ref{fig:loc}(b), we show a comparison of the ground state energies of the truncated and full chains and find an excellent agreement between the two results. This further confirms the mechanism of localisation we presented above.

Lastly, we briefly comment on the $\lambda>0$ case. 
The ground state localisation properties are similar to the case of $\lambda<0$, and the recursive relation underpinned by the hybridisation of two $L_{p-1}$ subsequences into the $L_p$ sequence, as well as the centre of $L_p$ coinciding with the mirror-symmetry point, hold in the same way. 
The only difference is that the localisation sequences $L_p$ are altered and no longer given by Tab.~\ref{tab:loc}. For example, the localisation in the first phase $p=1^+$ spreads over three sites that are connected by two degenerate hopping terms, see case (\textit{iv}) in Fig.~\ref{fig:hopping_modulation}(b). 
The origin of such a difference is in the fact that the strong and weak hopping terms exchange roles upon $\lambda \to -\lambda$ and, therefore, instances of two strong ($t>1$) nearest-neighbour hopping terms can occur in the chain for $\lambda>0$. 
This is confirmed by a simple analytical diagonalisation of a chain with three sites sharing two identical hopping terms that leads to a ground state with density distribution $(1/4, 1/2, 1/4)$ on each site, which gives the ${\rm IPR} = 3/8$ -- explaining why $\alpha_+ = 3/4$ in Eq.~\eqref{eq:IPR_GS} correctly maps the steps in the $\lambda>0$ curve to the ones in the $\lambda<0$ curve in Fig.~\ref{fig:phase_diagram_E0}(b).

%
\subsection{States in the middle of the spectrum    \label{subsec:middleband}}
%
Next, we turn our attention to the middle band, which hosts the states known as atomic states in the Fibonacci limit. As discussed in Sec.~\ref{sec:model}, these states share similar localisation properties with the ground state in the two limits; namely, they exhibit an extended-to-critical phase transition in the Aubry-Andr\'{e} limit, see Sec.~\ref{subsec:AA_limit}, while they are always critical in the Fibonacci limit, see Sec.~\ref{subsec:Fibonacci_limit}. Although critical in the latter limit, their fractal dimensions $D_q$ are almost equal to zero~\cite{Mace2016}, indicating that they are, in a sense, more localised than the states at the edges of the spectrum~\cite{Kohmoto1987}. 

%
\subsubsection{Localisation properties of the $E=0$ state  \label{subsec:E0_state}}
%
We start the analysis by considering a state in the middle of the spectrum, with $E = 0$, known to be of the Sutherland-Kalugin-Katz type~\cite{Kalugin2014,Mace2016, Mace_thesis2017}. The state is always critical and self-similar~\cite{Kohmoto1986,Kohmoto1987}. The $E=0$ state appears if the number of sites in an open chain is odd. Assuming that such a state exists for a system with $L=F_n$ number of bonds, or equivalently with $L+1$ number of sites, we can write the Schr\"{o}dinger equation for the spatial components of the $\psi(E=0)$ wavefunction. This gives a set of coupled equations, 
\begin{align}   \label{eq:schrodinger_e0}
    - t_j \psi_j(E=0) - t_{j+1} \psi_{j+2}(E=0) = 0 \, .
\end{align}
which for finite chains with open boundary conditions contain $t_0 \psi_1 = 0$ and $t_{L-1} \psi_{L-1} = 0$ from the chain's edges. Taking $L$ to be an even number, the two edge equations imply that the wavefunction $\psi_{j}$ has zero support on all sites with odd indices. For the remaining (even) sites, the wavefunction has the following form
\begin{align}   \label{eq:wavefunction_E0}
    \psi_{2l}(E=0) 
    &= \frac{\prod\limits_{m=0}^{l-1} (-1)^m \frac{t_{2m}}{t_{2m+1}}}{1 + \sum\limits_{l=1}^{\lfloor{L/2\rfloor}} \, \prod\limits_{m=0}^{l-1} \frac{t_{2m}^2}{t_{2m+1}^2}} \, ,
\end{align}
where we used Eq.~\eqref{eq:schrodinger_e0} and the normalisation condition.

\begin{figure}[t!]
    \centering
    \includegraphics[scale=1]{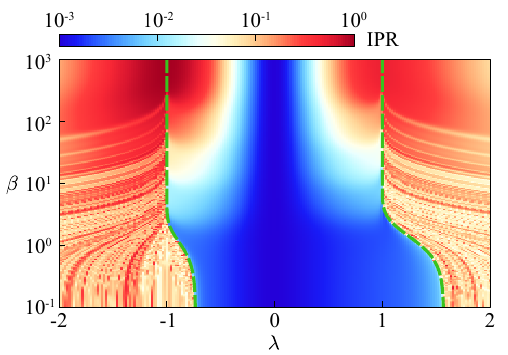}
    \caption{
    The localisation phase diagram of the $E=0$ state in Eq.~\eqref{eq:wavefunction_E0} obtained from the IPR as a function of $\lambda$ and $\beta$ for a system size $L=F_{14}=610$. Similarly to Fig.~\ref{fig:phase_diagram_GS}, the green dashed lines mark the analytical predictions for $\lambda^{\pm}_c$ given by Eq.~\eqref{eq:ldc}.
    }
    \label{fig:phase_diagram_E0}
\end{figure} 
From the $E=0$ eigenfunction, we can calculate the IPR and draw up the localisation phase diagram in Fig.~\ref{fig:phase_diagram_E0} as a function of $\lambda$ and $\beta$. For small $\beta$, the state is either extended or critical, depending on the value of $\lambda$, with the phase boundary given by Eq.~\eqref{eq:ldc}, as expected from the known Aubry-Andr\'{e} limit. When $\beta$ is increased, the IPR grows continuously without signatures of transitions between different localised phases, as we observed instead in the ground state. Moreover, while in the ground state the critical values of $\lambda$ lose their importance in favour of the cascade of localised phases when $\beta \gtrsim 1$, for the $E=0$ state they continue to bound the critical region even at high $\beta$. 

\begin{figure}[t!]
    \centering
    \includegraphics[scale=1]{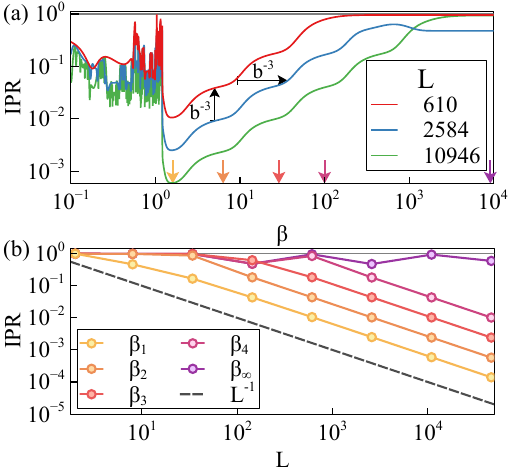}
    \caption{ 
    (a) IPR of the $E=0$ state in Eq.~\eqref{eq:wavefunction_E0} as a function of $\beta$ for three different sizes $F_{14}=610$, $F_{17}=2584$ and $F_{20}=10946$. Black vertical and horizontal arrows mark the scaling -- given by the coefficient $b^{-3}$ -- with length and $\beta$, respectively. The value of $\lambda$ in all three curves is $\lambda = -0.9$.
    (b) Scaling with the length for constant $\lambda = -0.9$ and for the values of $\beta$ marked by downward coloured arrows in (a). 
    }
    \label{fig:IPR_scaling_E0}
\end{figure}

To confirm the localisation behaviour observed in the phase diagram in Fig.~\ref{fig:phase_diagram_E0}, we plot in Fig.~\ref{fig:IPR_scaling_E0}(a) the IPR as a function of $\beta$ and at constant $\lambda=-0.9$ for three different system sizes. 
The behaviour of the three curves can be divided into three regions, depending on the value of $\beta$. At small $\beta$, i.e., for $\beta \lesssim 1$ for the choice of $\lambda$ in Fig.~\ref{fig:IPR_scaling_E0}(a), all three curves strongly fluctuate, which, together with the fact that larger systems have smaller IPR values in this region, identifies the critical phase (see also Fig.~\ref{fig:phase_diagram_E0}). 
For intermediate values of $\beta$ -- see App.~\ref{app:rg_1} for a more precise definition of the intermediate $\beta$ interval -- the curves exhibit periodic steps around an algebraic growth. Interestingly, in this interval the curves scale with length as $IPR \sim 1/L = 1/F_n$ -- where $F_n$ is the Fibonacci number -- unequivocally showing the presence of an extended phase. Consequently, if we denote the IPR of a chain of $F_n$ bonds as ${\rm IPR}^{(n)}$, then the scaling with length means that ${\rm IPR}^{(n)}(\beta) \approx b \, {\rm IPR}^{(n-1)}(\beta)$, where we used the fact that $b = \lim_{n \rightarrow \infty} F_{n-1}/F_{n}$. This explains the vertical scaling observed in Fig.~\ref{fig:IPR_scaling_E0}(a), which is further confirmed in Fig.~\ref{fig:IPR_scaling_E0}(b), for the choices of $\beta$ marked with colour-coded downward arrows in Fig.~\ref{fig:IPR_scaling_E0}(a).

Remarkably, the IPR scaling with $\beta$ with coefficient $b^{3}$ also works in the extended phase, as shown by the horizontal black arrow in Fig.~\ref{fig:IPR_scaling_E0}(a). This scaling only works for chains that have sizes which differ by a multiple of three in terms of Fibonacci numbers, or in other words 
\begin{align}   \label{eq:E0_IPR_scaling_with_beta}
    {\rm IPR}^{(n)}(\beta) \approx {\rm IPR}^{(n-3)}(b^{3} \, \beta) \, .
\end{align}
The scaling is exact in the $|\lambda| \rightarrow 1$ limit, as discussed in App.~\ref{app:scaling_of_psi}. 
Combining Eq.~\eqref{eq:E0_IPR_scaling_with_beta} with the vertical scaling, we obtain an equivalence relation for the IPR of each length separately, namely ${\rm IPR}^{(n)}(\beta) \approx b^{-3} {\rm IPR}^{(n)}(b^{3} \beta)$. This means that, for intermediate values of $\beta$, the shape of the steps in the IPR in Fig.~\ref{fig:IPR_scaling_E0}(a) is almost identical.

Lastly, in the regime of large $\beta$, the curves saturate, indicating the presence of a localised phase at least in finite-size systems. The saturation point shifts towards higher $\beta$ for larger systems, see App.~\ref{app:rg_1}. As we observe in Fig.~\ref{fig:IPR_scaling_E0}(a), this does not occur linearly with $L$, but follows a dependence that is intertwined with the atomic renormalisation discussed in Sec.~\ref{subsec:Fibonacci_limit}, as we explain in the analysis below. These observations are in agreement with previous knowledge about the $E=0$ state in the Fibonacci limit~\cite{Kohmoto1986,Piechon1995,Mace2016,Mace_thesis2017}.

%
\subsubsection{Renormalisation group approach at finite $\beta$}
%
The scaling of the IPR in Eq.~\eqref{eq:E0_IPR_scaling_with_beta} reminds of the atomic scaling in Sec.~\ref{subsec:Fibonacci_limit}, which relates chains of size $F_n$ to size $F_{n-3}$. This suggests a connection between $\beta$ and the atomic RG scheme, where $\beta$ continuously connects different RG steps. 
Intuitively, increasing $\beta$ has an effect on the $E=0$ state similar to the atomic RG scheme; namely, it reduces the spatial support of the fully extended wavefunction at small $\beta$ to a critical -- and almost localised -- behaviour for $\beta \rightarrow \infty$. Tuning $\beta$ and applying atomic RG steps keeps the middle state of the spectrum at $E=0$.

A crucial step towards understanding the connection between $\beta$ and atomic RG lies in the self-similarity of the hopping modulation function $V_j(\lambda, \beta)$. As detailed in App.~\ref{app:selfsimilarity}, the periodic function $V_j$ encodes a specific scaling relation for points located around its roots, which are given by $j_{1} = -1$ and $j_{2}=-2$ for the choice of phase $\varphi=3 \pi b$. Namely,  
$$
\tanh(b \beta) \cdot V_{j^{(n-1)}_{1/2}}(b \beta) \approx \tanh(\beta) \cdot V_{j^{(n)}_{2/1}}(\beta) \, ,
$$
where $j^{(n)}$ and $j^{(n-1)}$ belong to the chains with $L=F_n$ and $L=F_{n-1}$, respectively. More precisely, such scaling connects hopping modulation values $V_j$ around one of the two roots in a chain with $L=F_{n-1}$ to values around the other root in a chain with $L=F_n$.
As a consequence of such scaling, the hopping terms in the two chains are related by
\begin{align}   \label{eq:renormalised_hoppings}
    t_{j^{(n)}}(\lambda, \beta) \approx t_{j^{(n-1)}} \left( \tilde{\lambda}(\beta), b \beta \right) \, ,
\end{align}
with 
\begin{align}   \label{eq:tilde_lambda}
    \tilde{\lambda}(\beta) = \frac{\tanh(b \beta)}{\tanh(\beta)} \lambda \, .
\end{align}
For large $\beta$, the prefactor in Eq.~\eqref{eq:tilde_lambda} becomes unity and $\lim_{\beta \rightarrow \infty}\tilde{\lambda}(\beta) = \lambda$, so that Eq.~\eqref{eq:renormalised_hoppings} is mostly governed by the $\beta \rightarrow b \beta$ scaling, see Fig.~\ref{fig:selfsimilarity_hopping}.

Combining the scaling of the hopping terms in Eq.~\eqref{eq:renormalised_hoppings} with the closed analytical form of $\psi(E=0)$ in Eq.~\eqref{eq:wavefunction_E0} suggests the extension of the same scaling of $\beta$ in Eq.~\eqref{eq:renormalised_hoppings} to the wavefunction.
More precisely, the aforementioned wavefunction has a certain structure of maxima and minima depending on whether $t_{2m}/t_{2m+1}$ is a small or large quantity. 
As it can be seen in Fig.~\ref{fig:hopping_modulation}(b), a systematic alternation of weak and strong neighbouring hopping terms occurs around the symmetry point (\textit{iv}), namely around the position of the almost-degenerate hopping terms with $t_j \approx 1$. Consequently, the maxima and minima of $\psi(E=0)$ are concentrated around parts of the chain containing such $t_j \approx 1$ pairs. 
From the scaling of the $t_j \approx 1$ terms in Eq.~\eqref{eq:renormalised_hoppings}, we conclude that the scaled wavefunction $\psi^{(n-3)}(\tilde{\lambda}(\beta), b^3 \beta)$ has the same number of maxima as $\psi^{(n)}(\lambda, \beta)$, and they also share similar values, see App.~\ref{app:scaling_of_psi}.
The two aforementioned wavefunctions then have the same IPR, which explains the scaling in Eq.~\eqref{eq:E0_IPR_scaling_with_beta}.

%
\subsubsection{Other states near the middle of the spectrum}
%
%
\begin{figure}[t!]
    \centering
    \includegraphics[width=\columnwidth]{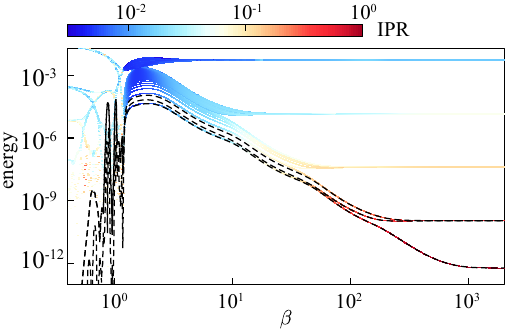}
    \caption{
    The behaviour of the IPR across the positive part of the spectrum with emphasis on the middle band. Note that the energy scale is given in the log scale. Dashed and solid black lines are analytical predictions given by the RG approach described in App.~\ref{app:rg_1}.
    We used a system with $L=F_{15}=987$ bonds and imposed periodic boundary conditions, while the strength of the hopping modulation is $\lambda = -0.9$. 
    }
    \label{fig:spec_middelband}
\end{figure} 
We next investigate the behaviour of other states in the middle band of the spectrum, see Figs.~\ref{fig:spectra}(b) and (c).
In Fig.~\ref{fig:spec_middelband}, we show the behaviour of the IPR across the middle band for $E > 0$, using a logarithmic scale to enhance the visibility of the behaviour of the states around $E=0$. 
At large $\beta$, the middle states bundle into subbands equidistant in energy and separated by $\rho^2 \equiv (t_w/t_s)^2 = \left( (1-\abs{\lambda})/(1+\abs{\lambda}) \right)^2$, as expected from the RG analysis in Sec.~\ref{subsec:Fibonacci_limit}, see Eq.~\eqref{eq:ham_rec_fib} and Refs.~\onlinecite{Niu1986,Niu1990,Piechon1995,Mace2016,Jagannathan2021}. A lower IPR for subbands higher in energy indicates that the states are more delocalised, in comparison to the states lower in energy. Again, this is in agreement with the Fibonacci model, where the approximate energy of each subband can be obtained by applying the atomic RG step to the bundle one step higher in energy, from which it follows that the number of states reduces for each subband lower in energy.
Furthermore, by reducing $\beta$, the energy of each bundle increases, starting from the lowest ones first, whose behaviour is influenced by the values of the hopping terms that are the closest to the middle of the distribution of $t_j$. The lower energy bundles then hybridise with the higher bundles as $\beta$ is further decreased, which gives the intuitive explanation of the steps observed in Fig.~\ref{fig:IPR_scaling_E0}(a). Additionally, the fact that the steps in IPR occur at $\beta$ where the lowest subband hybridises with the subband higher in energy, and knowing that subbands in the $\beta \rightarrow \infty$ limit are related by the atomic Fibonacci RG steps, hints towards the possibility of generalising the Fibonacci RG to finite $\beta$. Note also that this simple hybridisation procedure persists until $\beta$ is low enough to hit the critical region marked by the green dashed lines in Fig.~\ref{fig:phase_diagram_E0}, when all states in the middle band become critical. 

Motivated by the conclusions above, and knowing the relation in Eq.~\eqref{eq:renormalised_hoppings}, we develop the finite-$\beta$ RG formalism in App.~\ref{app:rg_1}.
From this new RG scheme, we obtain the energies of the first few states closest to the middle of the spectrum and we show them as dashed black lines in Fig.~\ref{fig:spec_middelband}. The agreement between these estimates obtained using finite-$\beta$ RG and the ones obtained numerically is very good at large $\beta$, with the discrepancy increasing with decreasing $\beta$. Surprisingly, the finite-$\beta$ RG nicely captures well the transition from critical to extended spectrum occurring around $\beta \approx 1$, even though this is a region where we do not expect the assumptions of the method to hold strongly.

%
\section{Discussion and conclusions  \label{sec:discussion_and_conclusions}}
%
The analysis of the interpolating off-diagonal IAAF model showed a plethora of interesting single-particle localisation properties, which we discuss in detail above. Let us now provide a summary from a broader perspective. 

It is clear from Fig.~\ref{fig:spectra} that the spectrum of the off-diagonal IAAF model can be divided into three main bands separated by the two largest gaps. The three bands are composed of two sets of states, namely molecular states, which symmetrically form upper and lower bands, and atomic states, which generate the band around zero energy -- following the convention in the Fibonacci limit.
Both molecular and atomic states have extended-to-critical phase transition as a function of $\lambda$ for small $\beta$, see Fig.~\ref{fig:spectra}(a), Fig.~\ref{fig:phase_diagram_GS}(a) and Fig.~\ref{fig:phase_diagram_E0}, whose boundary is given by Eq.~\eqref{eq:ldc}. However, their behaviour at larger, but finite, $\beta$ is drastically different. As it is shown in Fig.~\ref{fig:spectra}(c), starting from the critical phase and increasing $\beta$ whilst keeping $\lambda$ constant, atomic states undergo a critical to extended phase transition after which they remain extended for any finite $\beta$; the molecular states, in contrast, experience a critical to extended phase transition followed by multiple extended-to-localised transitions, i.e., multiple re-entrant localisation transitions~\cite{Strkalj2020}. 

To explain the behaviour of molecular states, we focused on the ground state and numerically studied higher moments of spatial eigenstate density~\eqref{eq:IPR}.
The phase diagram obtained using the IPR, finite-size scaling and fractal dimensions $D_q$ unequivocally shows the aforementioned cascade of re-entrant localisation-delocalisation transitions. We establish that the characteristic features of these re-entrant transitions and the mechanism responsible for their appearance are similar to the ones of the diagonal model studied in Ref.~\onlinecite{Strkalj2020}. However, in Sec.~\ref{subsec:gs}, we go a step further and discover a novel connection between the position of the localised ground state and the mirror-symmetry points of the underlying potential, see Fig.~\ref{fig:hopping_modulation}(b). The ground state localises on one of three symmetry points of the potential -- given by its minimum, maximum or the point where two neighbouring hopping terms with $t_j<1$ are almost degenerate -- in each localised region shown in Fig.~\ref{fig:phase_diagram_GS}(a), which alternate periodically. Moreover, we predict a finite segment over which the ground state density has the highest support in each localised region, and we summarise our findings in Table~\ref{tab:loc}. The results are further confirmed by the calculation of the ground state energy using the aforementioned finite localised sequence and the prediction of the spatial position of the localised ground state for different $\varphi$ in  App.~\ref{app:localisation_and_symmetry_points}.

The above mechanism of localisation does not hold for the atomic states in the middle band. Studying the $E=0$ state in Sec.~\ref{subsec:E0_state} revealed that no localisation is present at any value of $\beta$, as illustrated by the phase diagram in Fig.~\ref{fig:phase_diagram_E0} and the finite-size scaling of the IPR in Fig.~\ref{fig:IPR_scaling_E0}. This is radically different from the diagonal IAAF model discussed in Ref.~\onlinecite{Strkalj2020}, where all states in the spectrum showed re-entrant localisation transitions. 
For large systems, the $E=0$ state is always critical/extended for $|\lambda| \gtrless |\lambda_{\rm c}(\beta)|$, where $\lambda_{\rm c}(\beta)$ is given by Eq.~\eqref{eq:ldc}. 
Curiously, in the extended region, the IPR of the $E=0$ state shows a peculiar scaling given by Eq.~\eqref{eq:E0_IPR_scaling_with_beta}, which is a direct manifestation of the atomic RG scheme described in Sec.~\ref{subsec:Fibonacci_limit}. Specifically, once $\beta$ is tuned down from a very large -- essentially infinite -- value, the IPR decreases tracing a step-like structure, see Fig.~\ref{fig:IPR_scaling_E0}(a), with each step being connected to the neighbouring one with the same $b^{\pm 3}$ scaling as is the case in atomic RG.
In that sense, $\beta$ acts as a continuous variable for the aforementioned RG scheme. 

We explain the above scaling using the self-similarity of the modulation function~\eqref{eq:func_iaaf}, see App.~\ref{app:selfsimilarity}, combined with the scaling we derived for the hopping terms~\eqref{eq:renormalised_hoppings}.
Furthermore, motivated by the aforementioned scaling with $\beta$, we generalise the Fibonacci atomic RG scheme to a new finite-$\beta$ RG, see App.~\ref{app:rg_1}, and show that the energies of other states in the middle band, besides the $E=0$ state, can be reliably captured by it, see Fig.~\ref{fig:spec_middelband}.

The results presented in this work are of direct relevance to current experiments. For instance, the setup we discuss can be readily realised using cavity-polaritons~\cite{Strkalj2020, Milicevic2019, Pernet2022, Fontaine2022, Bloch2022}; the model we studied can be implemented as an array of coupled micropillars that host polaritons, with alternating centre-to-centre distances, similarly to Ref.~\onlinecite{Pernet2022}.
Another promising experimental platform are photonic lattices~\cite{Verbin2013,Verbin2015,Keil2011,Kremer2020,Kraus2012a}, where a plethora of off-diagonal models, including the IAAF model~\cite{Verbin2013,Verbin2015}, has already been realised.

Our results deepen the knowledge of complex phase transitions between extended, critical, and localised states, and enhance our theoretical understanding of quantum localisation phenomena in quasiperiodic systems. The relationship we presented between the localised ground state position and the system symmetry points opens new avenues for the manipulation and control of quantum states in engineered materials and devices. 
This could lead to the design of more efficient quantum sensors, novel quantum materials with tailored electronic properties, and advanced optical devices.

%
\section*{Acknowledgements}
%
We are grateful to Oded Zilberberg for fruitful discussions.
C.C. acknowledges financial support from the Engineering and Physical Sciences Research Council (EPSRC) grants No.~EP/T028580/1 and~EP/V062654/1.
The work of A.\v{S}. is supported by the European Union’s Horizon Europe research and innovation programme under the Marie Sk\l{}odowska-Curie Actions Grant agreement No. 101104378. 
All data that support the plots within this paper are available from the corresponding author upon request.

\appendix
%
\section{Ground-state localisation and symmetry points  \label{app:localisation_and_symmetry_points}}
%
\begin{figure}[h!]
    \centering
    \includegraphics[width=\columnwidth]{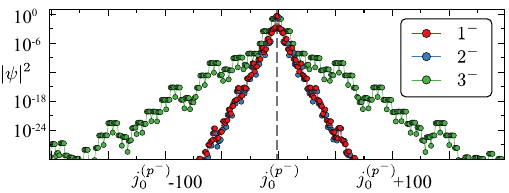}
    \caption{ 
    Density profile of the ground state wavefunction in the first three localised phases $p=1^-$ (red), $p=2^-$ (blue) and $p=3^-$ (green), where the latter two largely overlap (see also the discussion at the end of App.~\ref{app:localisation_and_symmetry_points}). A vertical dashed line marks the middle of the chain, i.e., $j_0^{(p^-)} = L/2$. Note that to obtain the same $j_0^{(p^-)}$ for different phases $p^-$, we used different $\varphi$ for each phase, as described in the text.
    The numerical parameters we used are $L=F_{16}=1597$ and $\lambda = -0.9$.
    }
    \label{fig:localisation_and_symmetry_points}
\end{figure}
To further confirm that the ground state indeed localises at the points of mirror symmetry discussed in Fig.~\ref{fig:hopping_modulation}(b), we find $\varphi$ in Eq.~\eqref{eq:func_iaaf} for each of the symmetry points (\textit{i}-\textit{iii}) that brings them to the middle of the chain, i.e., we impose $j_0^{(p^-)}=L/2$,
\begin{align}
    \varphi_{(i)} &= \pi - 2 \pi b \frac{L}{2} \, , \nonumber\\
    \varphi_{(ii)} &= - 2 \pi b \frac{L}{2} \, , \nonumber\\
    \varphi_{(iii)} &= - \pi b \left( 2 \frac{L}{2} - 1 \right) + \pi\, .
\end{align}
For example setting $\varphi = \varphi_{(i)}$ in Eq.~\eqref{eq:func_iaaf}, brings the symmetry point (\textit{i}) to $j_0^{(1^-)}=L/2$.
As we discussed in the main text, the localised phase $p=1^-$ is connected to (\textit{i}), $p=2^-$ to (\textit{ii}) and $p=3^-$ to (\textit{iii}). We numerically obtain the ground state density in the first three localised phases and for their respective $\varphi$ defined in the equations above, see Fig.~\ref{fig:localisation_and_symmetry_points}. We observe that all three curves share the same localisation point $j_0^{(p^-)}=L/2$, as predicted. 

Interestingly, we additionally notice in Fig.~\ref{fig:localisation_and_symmetry_points} that the wavefunctions in $p=1^-$ and $2^-$ phases have almost identical exponentially decaying tails. On the contrary, the tail of the wavefunction in a $p=3^-$ phase decays much slower. This indicates a complex dependence of the tail's decay rate as a function of $\beta$.

%
\section{Comparison of ground-state energies of truncated chains 
\label{app:GS_energy_comparison}}
%
\begin{figure}[t]
    \centering
    \includegraphics[width=\columnwidth]{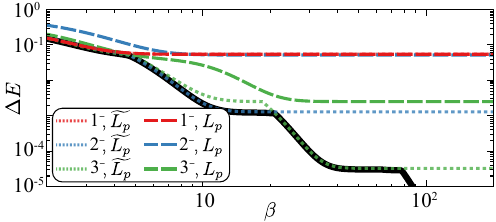}
    \caption{
    Similarly to Fig.~\ref{fig:loc}(b) in the main text, we show the ground state energy difference $\Delta E_{p^-} = E_{p^-} - E_{6^-}$ as a function of $\beta$. The black line is obtained by exact diagonalisation of a chain with $L = F_{17} = 2584$. Coloured dashed lines correspond to the energies $\Delta E_{p^{-}}$ obtained by exact diagonalisation of truncated chains with $L=L_{p}$. The coloured dotted lines are obtained for larger truncated chains with $L=\widetilde{L_p}$, as described in Eq.~\eqref{eq:larger_Lp} and also used in the main text.
    For all curves, we used the same hopping strength as in the main text, $\lambda=-0.9$.
    }
    \label{fig:GS_energy_comparison}
\end{figure}
Although most of the support of the ground state wavefunction is concentrated at the sites contained in $L_p$, given by Eq.~\eqref{eq:hoppings_loc_sequence}, the exponential tails are still crucial for obtaining the correct ground state energy after the truncation described in the main text. In other words, truncated chains containing only the $L_p$ sequence show a noticeable deviation from the ground state energy of the full system, as shown in Fig.~\ref{fig:GS_energy_comparison}. 
To make the agreement better, we consider a larger truncated chain with $L=\widetilde{L_p}$ bonds, which includes additional hopping terms around $L_p$, and it is given by
\begin{align} \label{eq:larger_Lp}
    t_j \in \widetilde{L_p} \:
    \mathrm{if} \: j \in \Big[& 
    \Big\lfloor j^{(p)}_0 - \frac{\delta_p+1}{2} \Big\rfloor
    -(F_{2p-3}+F_{2p-1}-1), \nonumber\\
    &\Big\lfloor j^{(p)}_0 + \frac{\delta_p+1}{2} \Big\rfloor
    +(F_{2p-3}+F_{2p-1}-1)
    \Big] \, .
\end{align}

Fig.~\ref{fig:GS_energy_comparison} shows the comparison of the ground state energy obtained for the two different types of truncation. Both the $L_p$ and the $\widetilde{L_p}$ truncated chains reproduce the correct ground state energy in the interval where the system is in the $p=1^-$ localised phase (see dotted and dashed red lines). However, in the region of $\beta$ where the system is in higher localised phases, $p \geq 2$, the correct energy is captured only by the $L=\widetilde{L_p}$ chain, as illustrated by the dotted lines.

%
\section{Self-similarity of the hopping terms   \label{app:selfsimilarity}}
%
\begin{figure}[h]
    \centering
    \includegraphics[width=\columnwidth]{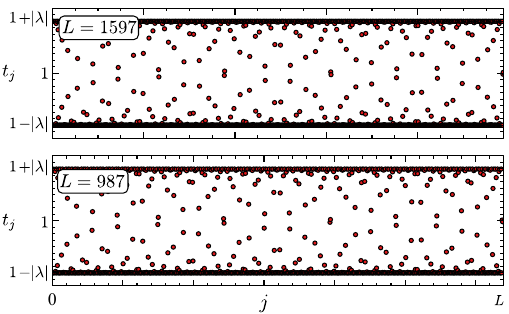}
    \caption{
    Self-similarity of the hopping terms between two different system sizes given by consecutive Fibonacci numbers $F_{15}=987$ (lower panel) and $F_{16}=1597$ (upper panel).   
    The value of $\lambda = -0.8$ is used for both plots, while $\beta=10$ in the upper panel is rescaled -- according to Eq.~\eqref{eq:renormalised_hoppings} -- to $b \beta \approx 6.18$ in the lower panel. 
    }
    \label{fig:selfsimilarity_hopping}
\end{figure}
In this Appendix, we derive the relation between the hopping terms in chains with different sizes given by two consecutive Fibonacci numbers, $F_n$ and $F_{n-1}$. The relation we obtain shows a universal scaling of $t_j$ with $\beta$, which we employ in the main text when discussing the extended phase -- occurring for intermediate values of $\beta$ (see App.~\ref{subsec:val_dom} for the definition of intermediate values) -- of the middle band in the energy spectrum. 

In the intermediate-$\beta$ regime, the values of $\beta$ strongly affect the hopping terms that are located in the middle of the distribution, around $t_j \approx 1$, see Fig.~\ref{fig:selfsimilarity_hopping}. On the contrary, `weak' and `strong' hopping terms that are located around $1+|\lambda|$ and $1-|\lambda|$ experience only minor corrections when $\beta$ is tuned. We shall therefore assume that the localisation properties of the middle band of states are mostly governed by the former type of hopping terms ($t_j \approx 1$), and in the following we concentrate solely on them.

%
\subsection{Main result   \label{subsec:result_selfsim}}
%
Before presenting a detailed proof, let us set the stage and present the main result of the scaling relation between the hopping terms with $t_j \approx 1$ for different system sizes. 

Firstly, let us introduce the function 
\begin{equation}    \label{eq:f}
    f(x) = -\cos(2 \pi b x + \varphi) + \cos(\pi b) \, , \quad {\rm with}\;  x \in \mathbb{R} \, .
\end{equation}
By taking a continuous version of Eq.~\eqref{eq:func_iaaf}, we can define the continuous hopping function $t(x,\lambda, \beta) \equiv 1 - \lambda \tanh \left[\beta f(x)\right] / \tanh(\beta)$.
The condition of $t_j \approx 1$ is then equivalent to
\begin{equation}    \label{eq:condition}
    \abs{f(j)} \ll 1 \, .
\end{equation}

Secondly, let us denote with $j^{(n)}$ the discrete indices $j$ -- representing the $j$-th bond and site of the chain -- that are part of the $L=F_n$ chain. 
The main result of this Appendix is that if we pick an index $j^{(n)} \in \left[ 0, F_{n} \right]$ (integer between $0$ and $F_{n}$) satisfying Eq.~\eqref{eq:condition}, we can find an index $j^{(n-1)} \in \left[ 0~,~F_{n-1} \right] $ also satisfying Eq.~\eqref{eq:condition}, such that
\begin{equation}    \label{eq:selfsim_gen}
     f(j^{(n-1)}) \simeq b^{-1} f(j^{(n)}) \, .
\end{equation}
The above leads to the final result for the hopping terms presented in Eq.~\eqref{eq:renormalised_hoppings} of the main text, 
\begin{align}  
    t_{j^{(n-1)}} \left(\widetilde{\lambda} (\beta), b \beta \right) \simeq  t_{j^{(n)}}(\lambda, \beta)\, , 
\end{align}
where the hopping strength scales as
\begin{align}
    \widetilde{\lambda}(\beta) = \frac{\tanh(b \beta)}{\tanh(\beta)} \lambda \, .
\end{align}
For $\beta \gg 1$, the prefactor tends to $1$ and we have $\widetilde{\lambda}(\beta) \approx \lambda$.

In Fig.~\ref{fig:selfsimilarity_hopping}, we show the spatial profile of hopping terms in chains with $L=F_{16}$ (upper panel) and $L=F_{15}$ (lower panel) for constant value of $\lambda$. In the larger chain we used $\beta=10$, while in the smaller chain $\beta$ is multiplied by the factor $b$, in accordance with Eq.~\eqref{eq:renormalised_hoppings}. We observe perfect agreement between the two panels of Fig.~\ref{fig:selfsimilarity_hopping} for the hopping strengths located between the two end values $1 \pm |\lambda|$.

%
\subsection{Proof and comments  \label{subsec:proof_selfsim}}
%
Let us now prove the main scaling result. To be consistent with the main text, we set $\varphi = 3 \pi b$, but the same approach can be used for other values of $\varphi$.

\subsubsection{Condition on indices $j^{(n)}$}
We use Eq.~\eqref{eq:condition} to impose the condition on the indices $j^{(n)}$. Firstly, the roots of the function $f(x)$ in Eq.~\eqref{eq:f} are
\begin{align}   \label{eq:exact_zero_f}
    f(x) = 0  \quad \iff \quad  x = -1 \; \; {\rm or} \, -2 \; (\bmod\; 1/b)
\, . 
\end{align}
for a choice of $\varphi$ we used in the main text.
The notation $\bmod\; 1/b$ refers to the modulo operation (here on real numbers). In the following, we concentrate on the $\lambda < 0$ case. Concentrating on a single period of the function $f(x)$, namely the interval $x \in \left[ \frac{-3}{2b}, \frac{-1}{2b} \right)$, it is easy to show that 
\begin{align}
    \begin{cases}
    f(x)<0 ; \qquad x \in (-2, -1) \\
    f(x)>0 ; \qquad x \in \left[ \frac{-3}{2b}, -2 \right) \cup \left( -1, \frac{-1}{2b} \right) \, .
    \end{cases}
\end{align}
Note that in the Fibonacci limit, a hopping term $t_j$ is weak (respectively, strong) when $f(j) < 0$ (respectively, $f(j)>0$). 

Since the function $f(x)$ has two zeros in a single period, we introduce two sets of indices defined as
\begin{equation}    \label{eq:set_middle}
    \mathcal{V}^{(n)}_i = \{\, j \in \left[ 0~,~F_{n} \right] \; \big| \; \exists \, \delta \ll 1\, ,\; j \equiv -i + \delta \;(\bmod\; 1/b) \} \, ,
\end{equation}
where $i = 1$ or $2$ are two zeros of $f(x)$, and $\delta$ is a small deviation from them. It follows that Eq.~\eqref{eq:set_middle} defines the points in the vicinity of the roots of $f(x)$. One can check that for $j^{(n)} \in \left[ 0, F_{n} \right]$
\begin{align} \label{eq:zero_f}
 \abs{f\left( j^{(n)} \right) } \ll 1 \quad  \iff \quad j^{(n)} \in \mathcal{V}^{(n)}_1 \cup \mathcal{V}^{(n)}_2 \, .
\end{align}
As a side remark, we note that the distance between two zeros of $f(x)$ is $1$, which means that the hopping terms in the middle of the distribution, i.e., $t_j\approx 1$, always come in pairs corresponding to two almost degenerate neighbouring hopping terms. 

\subsubsection{Relation between indices  \label{subsubsec:rel_indices}}
Let us now fix an integer $j^{(n)}_1 \in \mathcal{V}^{(n)}_1$ and find a related site index $j^{(n-1)}_i \in \mathcal{V}^{(n-1)}_i$ in the smaller chain. It is a priori not obvious whether $j^{(n-1)}$ will be in $\mathcal{V}^{(n-1)}_1$ or $\mathcal{V}^{(n-1)}_2$. However, in whichever interval $j^{(n-1)}$ lands, it has to respect the following three conditions
\begin{itemize}
    \item[ (*) ] $j^{(n-1)}_i$ is an integer ,
    \item[ (**) ] $0 \leq j^{(n-1)}_i \leq F_{n-1}$, 
    \item[ (***) ] $ j^{(n-1)}_i \equiv -i + \delta' \;(\bmod\; 1/b)$, with $\delta' \ll 1$ \, .
\end{itemize}

A naive guess for the relation between the two indices $j^{(n)}_1$ and $j^{(n-1)}_i$ would be a linear function with a coefficient $b$, since $F_{n-1} \simeq b F_{n}$ for large $n$. However, the irrationality of $b$ means that $b j^{(n)}_1$ is not an integer, and therefore it breaks condition (*). 

On the other hand, $\lfloor b j^{(n)}_1 \rfloor$ is an integer and such a term in $j^{(n-1)}_i$ would keep it integer. Therefore, using the definition
\begin{align}   \label{eq:define_k}
 \exists \delta \ll 1, \exists k \in \mathbb{Z}, \;  j^{(n)}_1 = -1 + \delta + k/b \, ,
\end{align}
we can compute $\lfloor b j^{(n)}_1 \rfloor$ to obtain
\begin{align} \label{eq:compute_floor}
    \lfloor b j^{(n)}_1 \rfloor &= \lfloor -b + \delta b \rfloor + k \nonumber \\
    &= -1 + k \\
    &= b (j^{(n)}_1 +1 -\delta)-1 \nonumber \, .
\end{align}
Now it is possible to define $j^{(n-1)}_i$ through
\begin{align}   \label{eq:define_jnm1}
    j^{(n-1)}_i = \lfloor b j^{(n)}_1 \rfloor - i +1 = b (j^{(n)}_1 +1 -\delta)-i \, .
\end{align}
Thanks to the first equality in Eq.~\eqref{eq:define_jnm1}, it follows that $j^{(n-1)}_i$ is an integer, and condition (*) holds. Condition (**) is also satisfied because $0 \leq j^{(n)}_1 \leq F_{n}$ and $F_{n-1} \simeq b F_{n}$. 
Lastly, we have to verify condition (***).
Using the property of the golden mean, i.e., $b = b^{-1} -1$, we write
\begin{align}   \label{eq:check3}
    j^{(n-1)}_i &= \left( \frac{1}{b} - 1 \right) (j^{(n)}_1 +1 -\delta)-i \nonumber \\
    &= \frac{1}{b}(j^{(n)}_1 +1 -\delta) - \frac{k}{b} -i \\
    &= -i -\frac{\delta}{b} + \frac{j^{(n)}_1 +1 -k}{b} \nonumber \, ,
\end{align}
where $j^{(n)}_1 + 1 - k$ is an integer. Therefore, we get to the conclusion
\begin{align} \label{eq:jnm1_congruence}
    j^{(n-1)}_i \equiv -i -\frac{\delta}{b} \; (\bmod \; 1/b) \, ,
\end{align}
which means that condition (***) is indeed satisfied. A similar relation can be found if we start from $j^{(n)}_2 \in \mathcal{V}^{(n)}_2$. 
Finally, to summarise, we showed that
\begin{align} \label{eq:index_relation}
    j^{(n)}_{i'} \in \mathcal{V}^{(n)}_{i'} \implies  j^{(n-1)}_i = b (j^{(n)}_1 +i' -\delta)-i \; \in \; \mathcal{V}^{(n-1)}_i \, ,
\end{align}
with $i, i' \in \{1, 2\}$. 

\subsubsection{Proof of the equality  \label{subsubsec:main_proof_selfsim}}
Knowing the relation between neighbourhoods around zeros of $f(x)$ for different chains, we continue with the derivation of Eq.~\eqref{eq:selfsim_gen}.
Let us fix $i = 1$ and $i' = 2$ in the expression given by Eq.~\eqref{eq:index_relation}. The value of $f$ is then
\begin{align}   \label{eq:compute_f}
    f\left( j^{(n-1)}_{1}\right) - \cos(\pi b) &= - \cos(2 \pi b^2 (j^{(n)}_{2} + 2 -\delta) + \pi b) \nonumber \\
    &= -\cos( 2 \pi b (j^{(n)}_{2}+2) + 2 \pi b^2 \delta - \pi b) \nonumber\\
    &= f\left( j^{(n)}_{2} + b\delta \right) - \cos(\pi b)  \, ,
\end{align}
where Eq.~\eqref{eq:index_relation} was used to derive the first and second line; we have also used the equation satisfied by the golden mean: $b^2 = 1-b$, together with  $2 \pi (j^{(n)}_{2} + 2 ) \in 2 \pi \mathbb{Z}$ and the parity of the cosine function. 

We now want to obtain the ratio $f(j_1^{(n-1)}) / f(j_2^{(n)})$, or equivalently, $f( j^{(n)}_{2} + b\delta ) / f(j_2^{(n)})$. Note that $\delta$ in Eq.~\eqref{eq:compute_f} is a small number, hence we are allowed to perform a Taylor series expansion
\begin{align}   \label{eq:taylor_f}
    \frac{f\left( j^{(n)}_{2} + b\delta \right)}{f\left( j^{(n)}_{2} \right)} \, \simeq \,  1 + b\delta \frac{1}{f\left( j^{(n)}_{2} \right)}.\frac{\partial f}{\partial j}\left( j^{(n)}_{2}\right) \, ,
\end{align}
retaining only the leading terms. 
Computing the denominator on the left-
\begin{align} \label{eq:compute_1derivatve_f}
    f \left( j^{(n)}_{2}\right) &= 2 \sin \left( \pi b j^{(n)}_{2} + 2\pi b \right) \sin \left( \pi b j^{(n)}_{2} + \pi b \right)  \nonumber \\
    &=  2 \sin \left( \pi b \delta \right) \sin \left( \pi b \delta -\pi b \right) \\
    &\simeq -2 \pi b \delta \sin \left( \pi b \right) \nonumber \, ,
\end{align}
and the first derivative on the right-hand side
\begin{align} \label{eq:compute_1derivatve_f}
    \frac{\partial f}{\partial j}\left( j^{(n)}_{2}\right) &= 2 \pi b \sin \left( 2 \pi b j^{(n)}_{2} + 3\pi b \right) \nonumber \\
    &= 2 \pi b \sin \left( 2 \pi b \delta - \pi b \right) \\
    &\simeq -2 \pi b \sin \left( \pi b \right) \nonumber \, ,
\end{align}
we arrive at
\begin{align} \label{eq:taylor_final}
    \frac{f\left( j^{(n)}_{2} + b\delta \right)}{f\left( j^{(n)}_{2} \right)} \, \simeq \, 1 +b \, = \, b^{-1} \, .
\end{align}
Finally, we can conclude that for $i \ne i'$
\begin{align}   \label{eq:conclusion}
    f \left( j^{(n-1)}_{i}\right) = b^{-1}  f \left( j^{(n)}_{i'}\right)\, .
\end{align}

%
\section{New atomic RG for finite $\beta$   \label{app:rg_1}}
%
To explain the main features of the middle band in the spectrum, which contains the atomic states in the extended phase, we develop a finite-$\beta$ renormalisation group scheme. The motivation comes from the Fibonacci limit ($\beta \rightarrow \infty$) of the model~\eqref{eq:ham}, where the atomic RG nicely captures the behaviour of the middle band when $|\lambda|$ is close to $1$. 
In what follows, we extend the atomic RG of the Fibonacci model by incorporating the $\beta$-dependence of the relevant hopping terms.

%
\subsection{Domain of validity  \label{subsec:val_dom}}
%
Let us start by properly defining the range of $\beta$ where the $E=0$ state in the middle band is extended. As observed in Fig.~\ref{fig:IPR_scaling_E0}, this range depends on the size of the chain, tending to infinity as $L$ is increased. 
We take $1-|\lambda| \ll 1$, for which a critical phase is present in the system for small $\beta$, i.e., for $\beta < \beta_{\rm c} \approx 1$ (see Fig.~\ref{fig:IPR_scaling_E0}). Increasing $\beta$ above $\beta_{\rm c}$ brings the system into an extended phase until it becomes critical (almost localised) for $\beta > \beta_f^{(n)}$. Hence, the intermediate $\beta$ values where the system is extended are given by 
\begin{align}
    \beta_{\rm c}(|\lambda| \approx 1) \approx 1 < \beta < \beta_f^{(n)}
\, . 
\end{align}

Recall that, in the Fibonacci limit, the hopping terms take only two values, $1 + \abs{\lambda}$ for strong and $1 - \abs{\lambda}$ for weak bonds. We say that our interpolating model is in the Fibonacci limit when every hopping term is $t_j \approx 1 \pm \abs{\lambda} \equiv 1 - \sign[f(j)] |\lambda|$. Such condition dictates that the interpolating function should be equal to $V_j(\beta) \approx \sign[f(j)]$, which is true if
\begin{align}   \label{eq:cond_fib_lim}
    \forall j \in \left[0, F_{n}\right] \quad \abs{\beta f(j)} \gg 1 \, .
\end{align}
From the above equation, we estimate $\beta^{(n)}_f$ for a chain with $F_n$ bonds
\begin{align}    \label{eq:beta_f}
     \beta^{(n)}_f \, \simeq \, \left[ \,\min\limits_{ j \in \left[ 0,F_{n} \right] }  \abs{f(j)} \quad \right]^{-1} \, .
\end{align}
In Tab.~\ref{tab:num_value_beta}, we calculate $\beta^{(n)}_f$ for a few system sizes and observe a good agreement with Fig.~\ref{fig:IPR_scaling_E0}.
\begin{table}[ht]
    \begin{tabular}{| c | c |} 
      \hline
      $F_n$ & $\beta^{(n)}_f$ \\ \hline 
      $610$ & $233.13$ \\ \hline
      $2584$ & $986.87$ \\ \hline
      $10946$ & $4179.76$ \\ \hline
    \end{tabular}
    \caption{Numerical values of $\beta^{(n)}_f$ for a few system sizes.} 
    \label{tab:num_value_beta}
\end{table}

%
\subsection{Defining the atomic sites for finite $\beta$ \label{subsec:at_sites_rg}}
%
We remind the reader that, in the Fibonacci limit, the atomic states in the middle band of a chain $L=F_n$ are captured by a smaller chain with $L=F_{n-3}$ that is obtained by performing a single atomic RG step on the initial chain. Then, by performing another atomic RG step, one obtains even smaller chain that captures the states in the middle subband of the middle band. Continuing to apply atomic RG steps will eventually lead to a short chain which captures only a few states around zero energy.  

The general algorithm for the atomic RG scheme is the following. As a first step, we need to identify the atomic sites, which is easily done in the Fibonacci limit, i.e. when $\beta > \beta_f$, by looking for sites surrounded by two weak hopping terms, see Fig.~\ref{fig:RG_Fib}(a). 
For $\beta < \beta_f$, we adopt the same procedure, but instead of weak hopping terms, we look for terms with values $t_j < 1$. Such terms become weak hopping terms in the $\beta \rightarrow \infty$ limit.
Once the atomic sites are identified, we need to calculate the effective hopping terms between them. From the Fibonacci sequence (see Sec.~\ref{sec:model}), it follows that the two pairs of letters $WW$ -- that have an atomic site in the middle -- can have either 3 or 5 letters between them. This condition is mapped onto the neighbourhood of the atomic sites, which can have either 3 or 5 bonds, i.e., hopping terms, between them. 

In App.~\ref{subsubsec:new_cluster}, to correctly account for the intermediate values of the hopping terms, we analyse clusters of 3 sites (or equivalently two bonds) that host a pair of neighbouring hopping terms which for finite $\beta$ strongly deviate from values $1 \pm |\lambda|$. More precisely, in this Subsection, we concentrate on one such pair that contains hopping terms that are furthest away from their values in the Fibonacci limit -- the one shown in (\textit{iv}) of Fig.~\ref{fig:hopping_modulation}(b) in the main text. In App.~\ref{subsubsec:neighbourhood_3site_cluster}, we explore the environment of these 3-site clusters and later apply Brillouin-Wigner perturbation theory in App.~\ref{subsec:BW_pert} to obtain the renormalised hopping terms after the finite-$\beta$ RG step. 

\subsubsection{3-site clusters \label{subsubsec:new_cluster}}
In App.~\ref{app:selfsimilarity}, as rigorously stated in Eq.~\eqref{eq:zero_f}, we found out that if a hopping term strongly deviates from its Fibonacci value at finite $\beta$, its nearest neighbour does also. In other words, when $\beta$ is decreased from $+ \infty$ to a finite value, the first hopping terms to migrate to intermediate values do so in pairs. More precisely, the first pair that starts deviating from the values $1 \pm |\lambda|$ is the almost degenerate pair shown in the symmetry point (\textit{iv}) in Fig.~\ref{fig:hopping_modulation}(b) in the main text.
Furthermore, we have seen that inside such a pair, one of the hopping terms comes from the strong hopping ($1+\abs{\lambda}$) while the other one comes from the weak one ($1-\abs{\lambda}$) in the Fibonacci limit. The two hopping terms in a pair have the following antisymmetric behaviour
\begin{align}   \label{eq:compute_f_j}
    f(j) \simeq -2 \pi b \delta \sin(\pi b) \simeq -f(j+1) \, ,
\end{align}
where we fixed $j \in \mathcal{V}_2$, (so $j+1 \in \mathcal{V}_1$) and used Eq.~\eqref{eq:compute_f}.

The hopping terms that surround the almost degenerate pair, i.e., $t_{j-1}$ and $t_{j+2}$, have
\begin{align}   \label{eq:compute_f_jp2}
    f(j-1) \simeq 2 \sin(\pi b)\sin(2 \pi b) \simeq f(j+2) \, ,
\end{align}
where $f(j-1)\simeq f(j+2) \simeq -1.3$ points out that the condition in Eq.~\eqref{eq:condition} is not satisfied, meaning that $t_{j-1}$ and $t_{j+2}$ are close to the values they have in the Fibonacci limit.
By recalling that $f$ is negative for these two hopping terms, which can be seen from (\textit{iv}) in Fig.~\ref{fig:hopping_modulation}(b) in the main text, it follows that
\begin{align}    \label{eq:compute_t_j}
    t_{j-1}(\lambda, \beta) \simeq t_{j+2}(\lambda, \beta) \simeq 1 - \abs{\lambda} \ll 1 \, .
\end{align}
In the Fibonacci limit, $t_{j-1}$ and $t_{j+2}$ become weak hopping terms, and one of $t_{j}$ and $t_{j+1}$ is also a weak hopping while the other is a strong hopping, see again (\textit{iv}) in Fig.~\ref{fig:hopping_modulation}(b) in the main text. 
Therefore, the 3-site cluster made of the almost degenerate hopping pair necessarily contains an atomic site.

\subsubsection{Neighbourhood of the 3-site clusters \label{subsubsec:neighbourhood_3site_cluster}}
Let us now investigate a larger neighbourhood of the aforementioned 3-site cluster that contains two additional atomic sites. Such neighbourhood contains 8 hopping terms, or equivalently 9 sites, and the middle atomic site is separated by 3 hopping terms from the atomic site on one end, and 5 from the site on the other end, see Fig.~\ref{fig:drawing_new_rg}.

From (\textit{iv}) in Fig.~\ref{fig:hopping_modulation}(b) in the main text, it is clear that if $t_{j} \approx t_{j+1}$, then $t_{j-3}, t_{j-1}, t_{j+2}, t_{j+4} < 1$ and $t_{j-2}, t_{j+3} > 1$ for any $\beta$. This can be explicitly proven in the same way as it was done in Eqs.~\eqref{eq:compute_f_jp2} and~\eqref{eq:compute_t_j}. Note that for any finite $\beta$, all hopping terms in the 8-hopping cluster are much closer to their Fibonacci values than the almost degenerate pair, namely $(t_{j-3}, t_{j-1}, t_{j+2}, t_{j+4}) \approx 1 - \abs{\lambda}$ and $(t_{j-2}, t_{j+3}) \approx 1 + \abs{\lambda}$. The former will therefore be treated as small perturbations in the next section.

%
\subsection{Brillouin Wigner perturbation theory \label{subsec:BW_pert}}
%
Now that we established the environment of the 8 hopping terms containing the almost degenerate hopping pair, we proceed with developing a new finite-$\beta$ atomic RG scheme. We show the 8-hopping sequence in Fig.~\ref{fig:drawing_new_rg}, where we denote the left hopping term in the almost degenerate pair as $t_{j_0}$ and the right one as $t_{j_0+1}$ -- this nomenclature follows from Eq.~\eqref{eq:exact_zero_f}. 

\subsubsection{Almost degenerate neighbouring hopping pair}
\begin{figure}[t]
    \centering
    \includegraphics[scale=1]{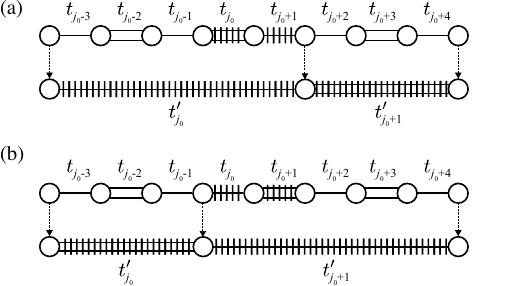}
    \caption{Two possible atomic renormalisation scenarios initially involving 8 hopping terms around the almost degenerate pair denoted by $(t_j,t_{j+1})=(t_{j_0},t_{j_0+1})$ and marked with vertical lines. A single (double) horizontal line denotes a weak (strong) hopping term with the value of $1-\abs{\lambda}$ ($1+\abs{\lambda}$).
    In (a) the left (right) hopping term in the pair is larger (smaller) than 1, i.e., $t_{j_0}>1$ ($t_{j_0+1}<1$), for any $\beta$, while in (b) the opposite is true. Note that other hopping terms in the sequence are the same in (a) and (b). Renormalised hopping terms are marked with $t_{j_0}'$ and $t_{j_0+1}'$.
    }
    \label{fig:drawing_new_rg}
\end{figure}
To derive an atomic renormalisation for a finite value of $\beta$, we need to compute effective hopping terms between two successive atomic sites. For this reason, we employ the Brillouin Wigner perturbation theory, with $\rho = (1-\abs{\lambda})/(1+\abs{\lambda})$ being a small parameter. Since at this point we concentrate on the case where $\beta$ is large and almost in the Fibonacci limit, all hopping terms in the 8-hopping sequence that have $t < 1$ will be considered small compared to the hopping terms that have $t > 1$, except the two hopping terms that are almost degenerate -- the ones marked with vertical lines in Fig.~\ref{fig:drawing_new_rg}.

We first focus on the sequence shown in Fig.~\ref{fig:drawing_new_rg}(a), where $t_{j_0}$ is infinitesimally larger than $t_{j_0+1}$. 
The Hamiltonian of such a sequence can be written as
\begin{align}   \label{eq:ham_pert_sum}
    H = H_0 + H_1 \, ,
\end{align}
with
\begin{align}   
    H_0 =& t_{j_0-2} | 1 \rangle \langle 2 | 
        + t_{j_0} | 3 \rangle \langle 4 | 
        + t_{j_0+1} | 4 \rangle \langle 5 |
        + t_{j_0+3}  | 6 \rangle \langle 7 | \nonumber \\  
        &+ h.c. \, , \nonumber \\  
    H_1 =& t_{j_0-3} | 0 \rangle \langle 1 | 
        + t_{j_0-1} | 2 \rangle \langle 3 | 
        + t_{j_0+2}  | 5 \rangle \langle 6 |
        + t_{j_0+4}  | 7 \rangle \langle 8 | \nonumber \\  
        &+ h.c. \, .
\end{align}
All hopping terms within $H_1$ are of order $1 - \abs{\lambda} \ll 1$. 
The atomic states of $H_0$, which have zero energy to first order in $\rho$, are given by:
\begin{align}    \label{eq:at_states}
    | \psi \rangle_0 &= | 0 \rangle  \nonumber\\
    | \psi \rangle_{\mu} &= \frac{t_{j_0+1} | 3 \rangle - t_{j_0} | 5 \rangle}{\mu}    \nonumber\\
    | \psi \rangle_8 &= | 8 \rangle \, ,
\end{align}
where we defined $\mu \equiv \sqrt{t_{j_0}^2 + t_{j_0+1}^2}$. 
On the other hand, the molecular states can be written as:
\begin{align}   \label{eq:mol_states}
        | \pm \rangle_1 &= \frac{| 1 \rangle \pm | 2 \rangle }{\sqrt{2}} \, ,   &&E = \pm t_{j_0-2} \nonumber\\
        | \pm \rangle_{\mu} &= \frac{| 4 \rangle}{\sqrt{2}} \pm  \frac{t_{j_0} | 3 \rangle + t_{j_0+1} | 5 \rangle}{\sqrt{2} \mu}  \, ,   &&E = \pm \mu \nonumber\\
        | \pm \rangle_6 &= \frac{| 6 \rangle \pm | 7 \rangle }{\sqrt{2}} \, ,   &&E = \pm t_{j_0+1} \, .
\end{align}

We can now define the projector $Q$ onto the atomic states 
\begin{align}    
    Q = | \psi \rangle_0 \langle \psi |_0
      + | \psi \rangle_{\mu} \langle \psi |_{\mu}
      + | \psi \rangle_8 \langle \psi |_8 \, , 
\end{align}
from which the effective Hamiltonian -- at any order $k \geq 0$ -- follows
\begin{align}   \label{eq:ham_eff}
    H^{\rm{eff}}_{0} &= Q H_0 Q \, ,\nonumber \\ 
    H^{\rm{eff}}_{k} &= H^{\rm{eff}}_{k-1} + Q H_1 \left( (\hat{\mathbf{1}} - Q) \frac{-1}{H_0}H_1 \right) ^{k-1} Q
    \, .
\end{align}

Having the effective Hamiltonian~\eqref{eq:ham_eff}, we can obtain the two renormalised hopping terms, see Fig.~\ref{fig:drawing_new_rg}(a), by calculating the expectation values up to the first non-vanishing order $k$: 
\begin{align}
    t_{j_0+1}' &= \langle \psi |_0 H^{\rm{eff}}_{k} | \psi \rangle_{\mu} = \frac{t_{j_0+2} t_{j_0+4}}{\mu \, t_{j_0+3}} t_{j_0}\, , \label{eq:tm1_prime}
    \\
    t_{j_0}' &= \langle \psi |_{\mu} H^{\rm{eff}}_{k} | \psi \rangle_8 = \frac{t_{j_0-3} t_{j_0-1}}{\mu \, t_{j_0-2}} t_{j_0+1} \, .
    \label{eq:tm2_prime}
\end{align}
The hopping terms in the almost degenerate pair exchange their positions after the atomic RG step; if $t_{j_0} > t_{j_0+1}$, then $t_{j_0}' < t_{j_0+1}'$.
Furthermore, it is easy to verify that in the Fibonacci limit, the expressions~\eqref{eq:tm1_prime} and~\eqref{eq:tm2_prime} give the known results: $t_{j_0+1}' = \rho^2 t_{s} \equiv t_{s}'$ and $t_{j_0}' = \rho^2 t_{w} \equiv t_{w}'$, see Sec.~\ref{subsec:Fibonacci_limit}.
As a consequence, the new finite-$\beta$ RG scheme is valid for both intermediate and large $\beta$ regimes, i.e., as long as the system is not in the critical region that appears for $\beta \leq \beta_{\rm c} \approx 1$, see Fig.~\ref{fig:IPR_scaling_E0}(a).

However, note an important difference between Eqs.~\eqref{eq:tm1_prime} and~\eqref{eq:tm2_prime}, and their counterparts in the Fibonacci limit. 
In the Fibonacci limit, $\mu = t_{j_0}$ in the case shown in Fig.~\ref{fig:drawing_new_rg}(a), meaning that a renormalised hopping term $t_{j_0+1}'$, which connects the second and the third atomic site in Fig.~\ref{fig:drawing_new_rg}(a), depends only on the hopping terms between those sites, i.e., $t_{j_0+1}' = t_{j_0+2} t_{j_0+4} / t_{j_0+3}$. On the other hand, for finite $\beta$, the hopping term $t_{j_0}$ -- located between the first and the second atomic site -- enters in the expression for $t_{j_0+1}'$. 

Lastly, by applying the Brillouin-Wigner perturbation theory to the situation drawn in Fig.~\ref{fig:drawing_new_rg}(b), one gets equations similar to Eqs.~\eqref{eq:tm1_prime} and~\eqref{eq:tm2_prime}, but with $t_{j_0+1}'$ and $t_{j_0}'$ exchanging places.

\subsubsection{Generalisation to all effective hopping terms}
\begin{figure}[t]
    \centering
    \includegraphics[width=\columnwidth]{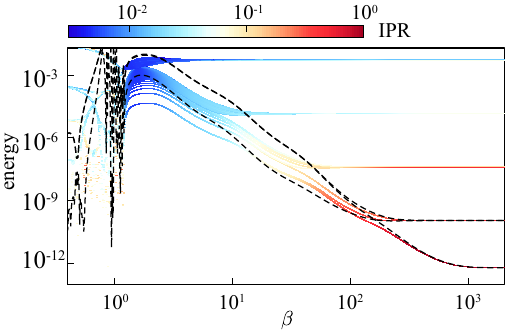}
    \caption{
    IPR across the positive part of the spectrum with emphasis on the middle band. Note the logarithmic energy scale. Dashed black lines are predictions given by Eqs.~\eqref{eq:tm2_prime_opt1} and~\eqref{eq:tm1_prime_opt1}.
    We used a system with $L=F_{15}=987$ bonds and imposed periodic boundary conditions, while the strength of the hopping modulation is $\lambda = -0.9$.
    }
    \label{fig:option1}
\end{figure}
Heretofore, we concentrated on the almost degenerate pair, which is well captured by the 8-hopping environment shown in Fig.~\ref{fig:drawing_new_rg}, and obtained renormalised hopping terms in Eqs.~\eqref{eq:tm1_prime} and ~\eqref{eq:tm2_prime}. However, there is a practical limitation to the RG scheme described above. Namely, one needs to first test whether the almost degenerate pair has $t_{j_0+1} > t_{j_0}$ or $t_{j_0+1} < t_{j_0}$, i.e., is the situation depicted by Fig.~\ref{fig:drawing_new_rg}(a) or (b), to correctly apply the equations for the effective hopping terms $t_{j_0+1}'$ and $t_{j_0}'$.

To resolve the aforementioned limitation and to be able to extend our considerations to the case of an effective hopping between any two atomic sites -- and not to only the ones containing the atomic site stemming from the almost degenerate hopping pair -- we generalise Eqs.~\eqref{eq:tm1_prime} and ~\eqref{eq:tm2_prime} by extending the 8-hopping sequence presented in Fig.~\ref{fig:drawing_new_rg} 
by a sequence containing one additional atomic site -- leading to a larger sequence with 11 or 13 hopping terms.
Hence, the renormalised sequence will contain 4 atomic sites separated by three effective (i.e., renormalised) hopping terms. In this new renormalised sequence of three hopping terms, we are interested only in the effective hopping term located in the middle. 

There are four possibilities for the number of hopping terms between 4 atomic sites, namely: 3-5-3, 5-3-5, 5-5-3, and 3-5-5 -- where the 3-5-3 denotes a sequence with 3 bonds between the first atomic site on the left end of the sequence and the second atomic site, 5 bonds between the second and the third atomic site and 3 bonds between the third and the fourth atomic site located at the right end of the sequence. Let us focus first on the aforementioned 3-5-3 situation, which is equivalent to Fig.~\ref{fig:drawing_new_rg}(a) but with 3 additional hopping terms on the left in the upper chain. Furthermore, we do not make assumptions on $t_{j_0-3}$ and $t_{j_0-2}$ being weak and strong hopping terms, as we did instead earlier. The expression for the middle effective hopping $t_{j_0}'$ follows from the Brillouin-Wigner perturbation theory, or simply by mathematical extension of Eq.~\eqref{eq:tm2_prime}, 
\begin{align}
    t_{j_0}' = \frac{t_{j_0-3}}{\sqrt{t_{j_0-3}^2 + t_{j_0-2}^2}} \; t_{j_0-1} \;  \frac{t_{j_0+1}}{\sqrt{t_{j_0}^2 + t_{j_0+1}^2}} \, .
    \label{eq:tm2_prime_opt1}
\end{align}
By setting $t_{j_0-2} \gg t_{j_0-3}$, we recover Eq.~\eqref{eq:tm2_prime}. 

If we focus on the case where the 3 hopping sequence is in the middle, i.e., the 5-3-5 case which corresponds to Fig.~\ref{fig:drawing_new_rg}(a) with additional 5 hopping terms on the right, we obtain the following
\begin{align}
     t_{j_0+1}' = \frac{t_{j_0}}{\sqrt{t_{j_0}^2 + t_{j_0+1}^2}} \; \frac{t_{j_0+2} t_{j_0+4}}{t_{j_0+3}} \; \frac{t_{j_0+6}}{\sqrt{t_{j_0+5}^2 + t_{j_0+6}^2}} \, , \label{eq:tm1_prime_opt1} 
\end{align}
where we do not make any assumptions on $t_{j_0+3}$ and $t_{j_0+4}$.
Setting $t_{j_0+6} \gg t_{j_0+5}$ recovers Eq.~\eqref{eq:tm1_prime}.

These refined expressions for $t_{j_0+1}'$ and $t_{j_0}'$ in Eqs.~\eqref{eq:tm2_prime_opt1} and~\eqref{eq:tm1_prime_opt1} allow us to obtain an effective hopping term between any two atomic sites in the chain.
The only requirement is the knowledge of the lengths of the hopping sequences between two atomic sites in the neighbourhood. In Fig.~\ref{fig:option1}, we show the numerical calculation of the first few energies closest to the middle of the band using the above expressions. The predictions, shown as dashed lines, agree well at large $\beta$, and start to deviate as $\beta$ is decreased. 

\begin{figure*}[t]
    \centering
    \includegraphics[scale=1]{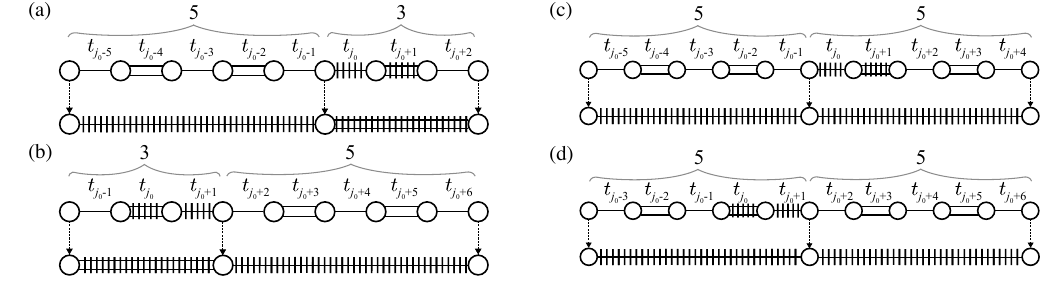}
    \caption{Four different configurations of hopping terms that can occur between three atomic sites. Numbers above the chains denote the number of bonds between two neighbouring atomic sites.
    }
    \label{fig:drawing_new_rg_generalised}
\end{figure*}

Let us push the argument used in the last few paragraphs further, in order to increase the agreement between the energies obtained by the RG approach and the energies obtained by exact diagonalisation of the whole chain. We calculate the effective hopping terms for four different situations occurring in the case of three atomic sites, shown in Fig.~\ref{fig:drawing_new_rg_generalised}, up to the second non-vanishing order in the Brillouin-Wigner perturbation theory and we combine them into the following expressions
\begin{align}
    t_{j_0}' =& \frac{t_{j_0-5}}{\sqrt{t_{j_0-5}^2 + t_{j_0-4}^2}} \; \frac{t_{j_0-3} t_{j_0-1} t_{j_0+1}}{\sqrt{t_{j_0-3}^2 + t_{j_0-2}^2} \sqrt{t_{j_0}^2 + t_{j_0+1}^2}} \nonumber\\ 
    & \times \frac{t_{j_0+3}}{\sqrt{ \displaystyle t_{j_0+2}^2 + t_{j_0+3}^2 } } \, ,
    \label{eq:tm2_prime_opt2}
\end{align}
for the case where the neighbouring atomic sites have 5 hopping terms between them, and
\begin{align}
     t_{j_0+1}' =& \frac{t_{j_0}}{\sqrt{t_{j_0}^2 + t_{j_0+1}^2}} \; \frac{t_{j_0+2} t_{j_0+4}}{\sqrt{\displaystyle t_{j_0+2}^2 + t_{j_0+3}^2 + t_{j_0+4}^2}} \nonumber\\ 
     & \times \frac{t_{j_0+6}}{\sqrt{\displaystyle t_{j_0+5}^2 + t_{j_0+6}^2}} \, , 
     \label{eq:tm1_prime_opt2} 
\end{align}
when there are 3 hopping terms between atomic sites.

Using the above expressions we calculate the dashed lines in Fig.~\ref{fig:spec_middelband}. The agreement with exact diagonalisation is better than in Fig.~\ref{fig:option1}. Additionally, these expressions for the renormalised hopping terms are more convenient since their form only depends on the length of the hopping sequence we renormalised; in other words, it depends only on whether there are 3 or 5 bonds between two neighbouring atomic sites.

%
\section{$b^3$ scaling of the $E=0$ wavefunction \label{app:scaling_of_psi}}
%
\begin{figure}[ht]
    \centering
    \includegraphics[width=\columnwidth]{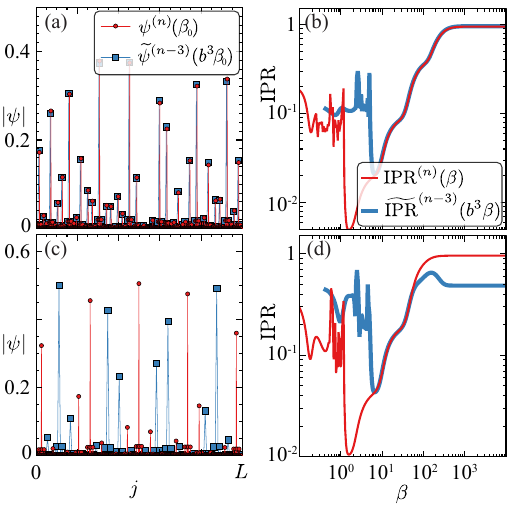}
    \caption{
    Finite-$\beta$ scaling for different boundary conditions. 
    (a) Plot of the absolute value of the wavefunction in the middle of the spectrum obtained by exact diagonalisation of chains with $L=F_n=987$ and $L=F_{n-3}=233$ and periodic boundary conditions. The absolute value of the same wavefunction for a smaller chain, $\psi^{(n-3)}$, is also shown multiplied by a prefactor $(1+\rho^2)^{-1}$ (see the text) and its argument $\beta$ scaled by $b^3$. 
    (b) The IPR as a function of $\beta$ for the state plotted in (a). The system sizes are the same as in (a), while the curve for the smaller systems is multiplied by a prefactor $(1+\rho^2)^{-4}$ with simultaneous scaling of $\beta$ by $b^3$.
    (c) and (d) show the same as (a) and (b) but for open boundary conditions. Here we used systems of size $L=F_n=610$ and $L=F_{n-3}=144$.
    For all plots, we keep $\lambda=-0.9$. In (a) and (c), we used $\beta_0=30$ for the chain with $L=F_n$, and $b^3 \beta_0 \approx 7.08$ for the chain with $L=F_{n-3}$. Both values lie in the region of the phase space where the system is extended.
    }
    \label{fig:scaling_of_E0_wavefunction}
\end{figure}
In Fig.~\ref{fig:scaling_of_E0_wavefunction} we show the scaling of the wavefunction in the middle of the spectrum and its IPR for both periodic and open boundary conditions. 
In Figs.~\ref{fig:scaling_of_E0_wavefunction}(a) and~(c), the wavefunction of the smaller chain with $L=F_{n-3}$ is scaled by the prefactor drawn from the atomic RG scaling in the Fibonacci model, see Refs.~\onlinecite{Mace2016, Mace_thesis2017},
\begin{align}
    \widetilde{\psi}^{(n-3)} (\beta) = \frac{1}{1+\rho^2} \;
 \psi^{(n-3)}(\beta) \, ,
\end{align}
with $\rho = (1-|\lambda|)/(1+|\lambda|)$ being the ratio of weak and strong hopping terms in the Fibonacci limit. Further, in the smaller chain, we also scaled $\beta$ by the factor $b^3$, as discussed in Sec.~\ref{subsec:middleband}.
The IPRs in Figs.~\ref{fig:scaling_of_E0_wavefunction}(b) and~(d) are obtained from $\psi^{(n)}(\beta)$ and $\widetilde{\psi}^{(n-3)}(b^3 \beta)$ using Eq.~\eqref{eq:IPR}.

For periodic boundary conditions, the maxima of the wavefunctions $\psi^{(n)}(\beta)$ and $\widetilde{\psi}^{(n-3)}(b^3 \beta)$ strongly overlap, which then gives overlapping curves in the IPR for different chains. 
Surprisingly, although the scaled $\widetilde{\psi}(b^3 \beta)$ does not match $\psi^{(n)}(\beta)$ in the case of open boundary conditions shown in Fig.~\ref{fig:scaling_of_E0_wavefunction}(c), their local maxima are similar in height and number, resulting in identical IPR in Fig.~\ref{fig:scaling_of_E0_wavefunction}(d) for the regime of $\beta$ discussed in App.~\ref{subsec:val_dom}. 

Lastly, note that the prefactor in front of the $\widetilde{{\rm IPR}}^{(n-3)} = \sum_j |\widetilde{\psi_j}^{(n-3)}|^4$ is almost unity: $1/(1+\rho^2)^4 \simeq 0.989$ for $\lambda=0.9$. Thus, the scaling of IPR between $L=F_n$ and $L=F_{n-3}$ chains is governed almost entirely by the $b^3$ scaling of $\beta$.

\clearpage
\newpage

%
%
%

\end{document}